\documentclass[useAMS,usenatbib]{mn2e}
\usepackage{epsfig,amsmath,amssymb,amsfonts,mathrsfs,latexsym,graphicx}
\begin{document}

\title[Hydrogen and helium in SNe IIb] {Hydrogen and helium in
  the late phase of SNe of Type IIb}

\author[Maurer et al.]{I. Maurer$^{1,**}$, P. A. Mazzali$^{1,2,3}$,
  S. Taubenberger$^{1}$, S. Hachinger$^{1}$
\\$^1$ Max Planck Institut f\"ur
  Astrophysik, Karl-Schwarzschild-Str.1, 85741 Garching, Germany
\\$^2$ Scuola Normale Superiore, Piazza dei Cavalieri, 7, 56126 Pisa, Italy
\\$^3$ National Institute for Astrophysics-OAPd, Vicolo dell'Osservatorio, 5, 35122 Padova, Italy
\\
\\$^{**} maurer@mpa$-$garching.mpg.de$
}

\maketitle

\begin{abstract}
Supernovae of Type IIb contain large fractions of helium and traces of
hydrogen, which
can be observed in the early and late spectra. Estimates of the hydrogen and
helium mass and distribution are mainly based on early-time spectroscopy and are
uncertain since the respective lines are usually observed in
absorption. Constraining the mass and distribution of H and He is important to gain insight
into the progenitor systems of these SNe.

We implement a NLTE treatment of hydrogen and helium in a three-dimensional
nebular code. Ionisation, recombination,
(non-)thermal electron excitation and H$\alpha$ line scattering are
taken into account to compute the formation of H$\alpha$, which is by far the strongest H
line observed in the nebular spectra of SNe IIb. Other
lines of H and He are also computed but are rarely identified
in the nebular phase. Nebular models are computed for the Type IIb SNe 1993J, 2001ig, 2003bg and
2008ax as well as for SN 2007Y, which shows H$\alpha$
absorption features at early times and strong
H$\alpha$ emission in its late phase, but has been classified as a SN
Ib. We suggest to classify SN 2007Y as a SN IIb. Optical
spectra exist for all SNe of our sample, and there is one IR nebular observation of
SN 2008ax, which allows an exploration of its helium
mass and distribution. 

We develop a three-dimensional model for SN 2008ax. We obtain estimates for the total mass and kinetic energy in good agreement
with the results from
light-curve modelling found in the literature. We further derive abundances of He, C, O, Ca
and $^{56}$Ni. Estimates of
the H mass are difficult but some constraints are derived. We demonstrate that
H$\alpha$ absorption is probably responsible for the double-peaked profile
of the [O {\sc i}] $\lambda\lambda$ 6300, 6363 doublet in
several SNe IIb and present a mechanism alternative to shock
interaction for generating late-time H$\alpha$ emission of SNe IIb.

\end{abstract}


\maketitle

\section{Introduction}
\label{intro}

Massive stars ($>8$ M$_\odot$) collapse when the nuclear fuel in their
central regions is consumed, producing a core-collapse supernova (CC-SN) and
forming a black hole or a neutron star. CC-SNe with a H-rich spectrum are classified 
as Type II. If the envelope was stripped to some degree prior to the explosion, the
SNe are classified as Type IIb (strong He lines, and weak but clear H), Type Ib (strong
He lines but no H), and Type Ic (no He or H lines) \citep{Filippenko97}.

Large amounts of
hydrogen ($\gg$ 1 M$\odot$) cause the progenitor to become a red
super-giant prior to explosion, which in turn leads to a broad peaked SN
light curve \citep[e.g.][]{Grassberg71}. Since this is not observed in SNe of Type
IIb there is an upper limit for the hydrogen mass of less than one
solar mass \citep{Nomoto93,Utrobin94,Woosley94}. It is not clear how SNe IIb manage to keep just a thin layer of hydrogen.

While massive (and therefore hot, O-Type) stars (M $>$ 25 M$_\odot$) can blow off their hydrogen
envelope by radiatively driven winds \citep[e.g.][]{Eldridge06},
mass estimates for the progenitors of SNe IIb \citep[e.g.][]{Woosley94,Mazzali09,Silverman09,Hamuy09} suggest that they may not be massive enough to lose most of
the H envelope via this process. Binary
interaction can cause mass transfer between two interacting stars and
would allow a Type IIb progenitor to lose most of its hydrogen
\citep[e.g.][]{Woosley94}. There is observational evidence that the
progenitors of SNe 1993J \citep{Aldering94,Maund04,Maund09}, 2001ig
\citep{Ryder06} [also see \citet{Kotak06} for an alternative
  interpretation of the SN 2001ig data]
and 2008ax \citep{Crockett08} may have been part of binary
systems. 

There is evidence of shock interaction from nebular H$\alpha$
\citep[e.g.][also see this work]{Chugai91,Patat95,Houck96}
as well as from radio
\citep[e.g.][]{Fransson98,Soderberg06,Chevalier10} and X-ray observations \citep[e.g.][]{Chevalier81,Soderberg06,Nymark09,Chevalier10}, suggesting that at
least some SNe IIb are surrounded by massive stellar
winds.
This wind could also be produced by the massive companion star in the
binary scenario.

At early
phases of SNe IIb the energy emitted by H$\alpha$ is probably provided
by the radioactive
decay chain $^{56}$Ni~$\to$~$^{56}$Co~$\to$~$^{56}$Fe. However, it has been
shown for SN 1993J \citep[][]{Patat95,Houck96} that $\sim$~150 days after explosion this
mechanism becomes ineffective suggesting that an additional source of energy
is required to explain the observed H$\alpha$
luminosities (however this may not be true; see Section \ref{clumping}). A similar result had been obtained
for SNe~II \citep{Chugai91}. This is further confirmed by the detection of a flattening of
the H$\alpha$ luminosity decay at late phases \citep{Matheson00}.
This could be inconsistent
with radioactive decay, but may be explained by shock interaction.

Shock interaction can also be
detected through radio and X-ray observations. As the SN ejecta propagate
into the circumstellar medium, they are decelerated, creating internal
energy which is dissipated by radiative processes. There are
several solutions for different scenarios
of shock interaction \citep[e.g.][]{Chevalier81,Suzuki95,Fransson98}, but a clear
interpretation of the observations is often
difficult. Micro-physical processes (e.g. the formation of magnetic
fields) are poorly understood, and it is not clear in detail how the energy released by
the shock is transferred into radiation. The SN envelope and
circumstellar density profiles are not known and have to be treated
as free parameters. Several authors found evidence for
inhomogeneous wind structures and deviations from wind-like
density profiles of the external medium
\citep[e.g.][]{Fransson94,vanDyk94,Suzuki95}. This may however be an artefact of an inaccurate treatment of shock
physics \citep{Fransson98,Fransson05}. A very careful
treatment of shock interaction seems necessary to obtain reliable
results \citep[e.g.][]{Fransson98}.

Asphericities in the inner and outer ejecta 
are evident in at least some CC-SNe. Three indicators are
velocity differences of Fe and lighter element lines at late times \citep[e.g.][]{Mazzali01}, polarisation
measurements \citep[e.g.][]{Hoflich91} and double-peaked or
asymmetric emission line profiles
\citep[e.g.][]{Mazzali05,Maeda08,Modjaz08,Taubenberger09,Maurer09}[also
  see \citet{Mili10} for an alternative interpretation, which is
  however doubtful \citep{Maurer09}]. Indirect
evidence also emerges from a comparison of the inner and outer
ejecta velocities \citep{Maurer09}. We find evidence
that SN 2008ax is a non-spherical event (see Section \ref{sn08ax}).

Between 100 and 200 days after explosion most SNe enter their nebular
phase. Owing to the decrease of density the SN ejecta become
transparent at most wavelengths, which makes it possible to observe the
innermost parts of the SN. Nebular modelling provides the opportunity to derive information about
the SN ejecta mass, kinetic energy, abundances and geometry, which can
hardly be obtained by other methods at least for the central regions.  

Nebular observations are available for very few SNe IIb. To our knowledge there are nebular spectra of SNe 1993J, 2001ig,
2003bg, 2006T, 2008aq and 2008ax. The quality of the SN 2006T spectrum is
poor. For SN 2008aq there is no light curve and therefore there
are no flux-calibrated spectra. Therefore, neither SNe are included in our
analysis.

Although early time H$\alpha$ absorption features are observed, SN 2007Y was classified as a SN Ib by \citet{Stritzinger09}, who argued
there is very little hydrogen in the envelope and that the
late H$\alpha$ emission is powered by shock interaction. On the other hand, \citet{Chevalier10} found that the
H$\alpha$ emission of SN 2007Y must be powered by radioactive decay
before 300 days after explosion. SN 2007Y is included in our analysis to
investigate this contradiction and a possible re-classification.

In Section 2 we describe the implementation of hydrogen and helium in
our nebular code. In Section 3 we
describe a three-dimensional model of SN 2008ax, give some
estimates for the total mass and kinetic energy and compare them to
results available in the literature. In Section 4 we present nebular models for
the SNe 1993J, 2001ig, 2003bg, 2007Y.  Since nebular models for these SNe are
available in the literature already and since we cannot determine their helium
density we concentrate on their H$\alpha$ emission. In Section \ref{clumping} we
present a scenario alternative to shock interaction for explaining
strong late-phase H$\alpha$ emission. In Section \ref{disc} our
results are discussed.

\section{H and He in the nebular phase}
In this section we describe the implementation of hydrogen and helium
in our nebular code \citep[][]{Ruiz92,Mazzali01,Mazzali07,Maurer09}. The code
computes the energy deposition from the radioactive decay of
$^{56}$Ni and $^{56}$Co in the SN ejecta, uses the energy deposited to
compute ionisation and excitation, and balances this with gas cooling
via line emission in a NLTE scheme. The method follows that developed
by \citet{Axelrod80}. 

\subsection{Hydrogen}

A full NLTE treatment of hydrogen is implemented in our
three-dimensional nebular code.
We obtain radiative transition and ionisation rates for the hydrogen
atom from analytical
solutions available in the literature \citep[e.g.][]{Burgess65}. Collisional rates
for allowed transitions between excited levels n $> 3$ are unimportant, but are included using an
approximation \citep{VanRegemorter62}. Collisional rates
between quantum levels n = 1,2 and 3 are taken from \citet{Scholz90,Callaway94}. 
The 2s level is additionally connected to the ground state by
two-photon decay, and the 2s and 2p
levels are coupled by electron and proton collisions. 

Hydrogen is ionised by non-thermal electrons (produced by Compton
scattering of $^{56}$Co radioactive decay $\gamma$-radiation) and by
UV-radiation emitted by helium if hydrogen and helium are mixed. The
electron impact ionisation rate can be obtained comparing the atomic and
electronic loss-functions with the ionisation cross-sections
\citep[e.g.][]{Axelrod80,Maurer10}. Thermal electron collisional
ionisation and recombination is not important.
Excitation by non-thermal electrons is important. We
compute non-thermal electron excitation rates using the
approximation of \citet{Rozsnyai80}.  

We compare our hydrogen and helium ionisation and non-thermal
excitation rates to calculations from Hachinger et. al. (in prep.) who
calculate non-thermal
ionisation and excitation rates solving an energy balance equation
derived from the Spencer-Fano equation \citep{Xu91,Lucy91}. Although these methods are quite
different, the thermal excitation and ionisation
rates at various electron
and atomic densities agree to 10\% $-$ 20\% for both hydrogen and helium.  

For levels with principal quantum number n $ > 5$ we only consider
recombination (these levels are almost completely depopulated and
have no influence on the nebular spectrum), while a full NLTE
(de-)excitation treatment is performed for lower levels.

The n = 2 level of H can effectively scatter the background radiation field. This is
a very important process in SNe IIb 
\citep[][]{Houck96}[Hou96]
since H$\alpha$ comes into resonance with the [O {\sc i}]
$\lambda\lambda$ 6300, 6363 doublet, which carries an important fraction
of the nebular flux. The optical depth of transitions between levels n$\ge$ 3 is too low to cause
observable line scattering at late epochs. 

Clumping of hydrogen is usually neglected in SNe IIb \citep{Patat95,Houck96,Mazzali09}, but it can increase the optical
depth of H$\alpha$ considerably. Under certain conditions, as we
show in Section \ref{clumping}, it can also increase the H$\alpha$ emission
significantly.

\subsection{Helium}
Accurate atomic data for He are available in the literature (e.g.
see TIPTOPbase \footnote{http://cdsweb.u-strasbg.fr/topbase/}, NIST \footnote{http://www.nist.gov/index.html}).
For technical reasons it was more convenient for us to calculate the He radiative transition and ionisation rates
using a quantum defect method \citep{Bates49,Seaton58,Burgess60}[also
  see \citet{Maurer10} for an application to oxygen], which is very accurate for helium. The disagreement between the calculations
and NIST recommended data is a few percent in the worst case, having no observable influence on the line calculations. 

Thermal electron excitation rates for all quantum levels n $<$ 4
are taken from \citet{Berrington87}. The 2s($^1$S) level is connected to the
ground state by two-photon emission (2PE) \citep[e.g.][]{Drake69}, which turns out to be
important in the nebular phase of SNe of Type IIb (in the early phase radiative excitation of
the 2p levels allows effective de-excitation via the 2p $\to$ 1s transition,
which can be more important than 2PE then).

Helium is mainly ionised by non-thermal electrons (see hydrogen). The
recombination rates into all levels n $\le$ 4 are calculated by
simulating recombination into and cascading from excited levels n
$\le$ 30. Levels with n $\le$ 4 are treated in full NLTE. Non-thermal
electron excitation is important and is
included using the approximation of \citet{Rozsnyai80}.

Helium can be observed in emission during the nebular phase of SNe IIb since its
mass is large and since it is distributed down to
a few thousand km s$^{-1}$. Owing to the large excitation potential of
He ($\sim$ 20 eV), an important fraction of the deposited energy can be radiated away
by other elements which may be present in small fractions. This means
that the electron temperature decreases and the He lines become weaker while the other elements emit more
strongly. Since all optical He lines are blended with lines only the IR
He {\sc i} 10830 \AA\ and 20587 \AA\ lines can be identified clearly
(at least in SN 2008ax), which allows an estimate of the He mass and
distribution. Since both lines are optically thick around 100 days
after explosion, they can be influenced by line scattering.

\section{SN 2008ax}
\label{sn08ax}

\begin{figure} 
\begin{center}
\includegraphics[width=8.5cm, height=8cm]{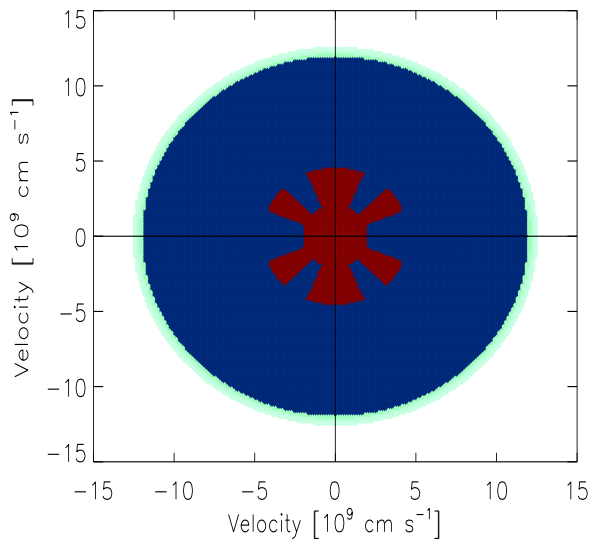}
\end{center}
\caption{Illustration of the three-dimensional model of SN
  2008ax, described in the text. Regions drawn in red colour are
  rich in heavy elements and contain no helium. Regions indicated in blue are helium-rich. The outermost region, indicated in green, is hydrogen-dominated. The x-axis is associated with the equator. In the centre (below 4500 km s$^{-1}$), heavy elements
  (red) are concentrated
  towards the pole, while ejecta containing helium (blue) are
  concentrated in the equatorial regions.}
\label{pap5sn08axpic}
\end{figure}

\begin{figure} 
\begin{center}
\includegraphics[width=8.5cm, clip]{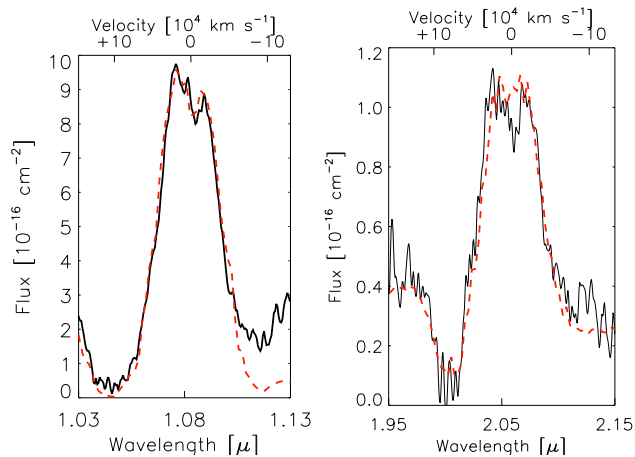}
\end{center}
\caption{IR spectrum of SN 2008ax at 131 days after explosion. The
  He {\sc i} 10830 \AA\ line is shown on the left and the He {\sc i}
  20587 \AA\ line is shown on the right. The observations are
  shown in black while the synthetic flux is shown by the dashed red
  line.  The synthetic flux is
  obtained using the three-dimensional model described in the
  text. Since almost no background flux is produced by our nebular code
  near 20000 \AA\ we added some constant flux at the level of the
  observations between 19000 and
  22000 \AA\ to allow He {\sc i} 20587 \AA\ line scattering. In contrast to the optical He
lines, the 10830 \AA\ and 20587 \AA\ lines can be identified clearly,
which allows a determination of the He density field. Two kinds of
asymmetry can be observed. First, the double-peaked nature suggests a
torus-shaped distribution of He. Second, the blue side of the profiles
of both lines
is stronger, which suggests an asymmetry along the line of sight.}
\label{pap5sn08ax131}
\end{figure}

\begin{figure} 
\begin{center}
\includegraphics[width=8.5cm, clip]{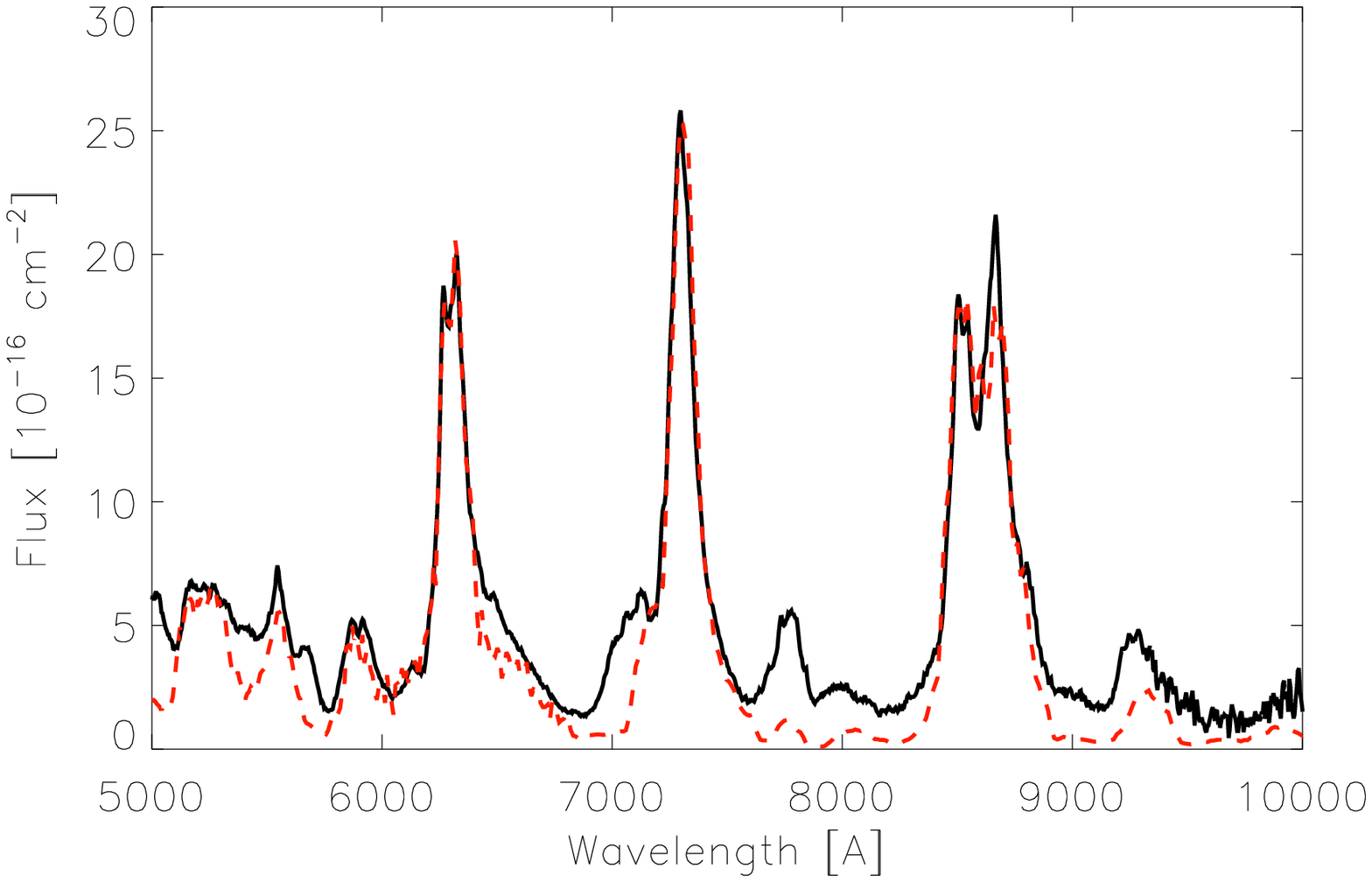}
\includegraphics[width=8.5cm, clip]{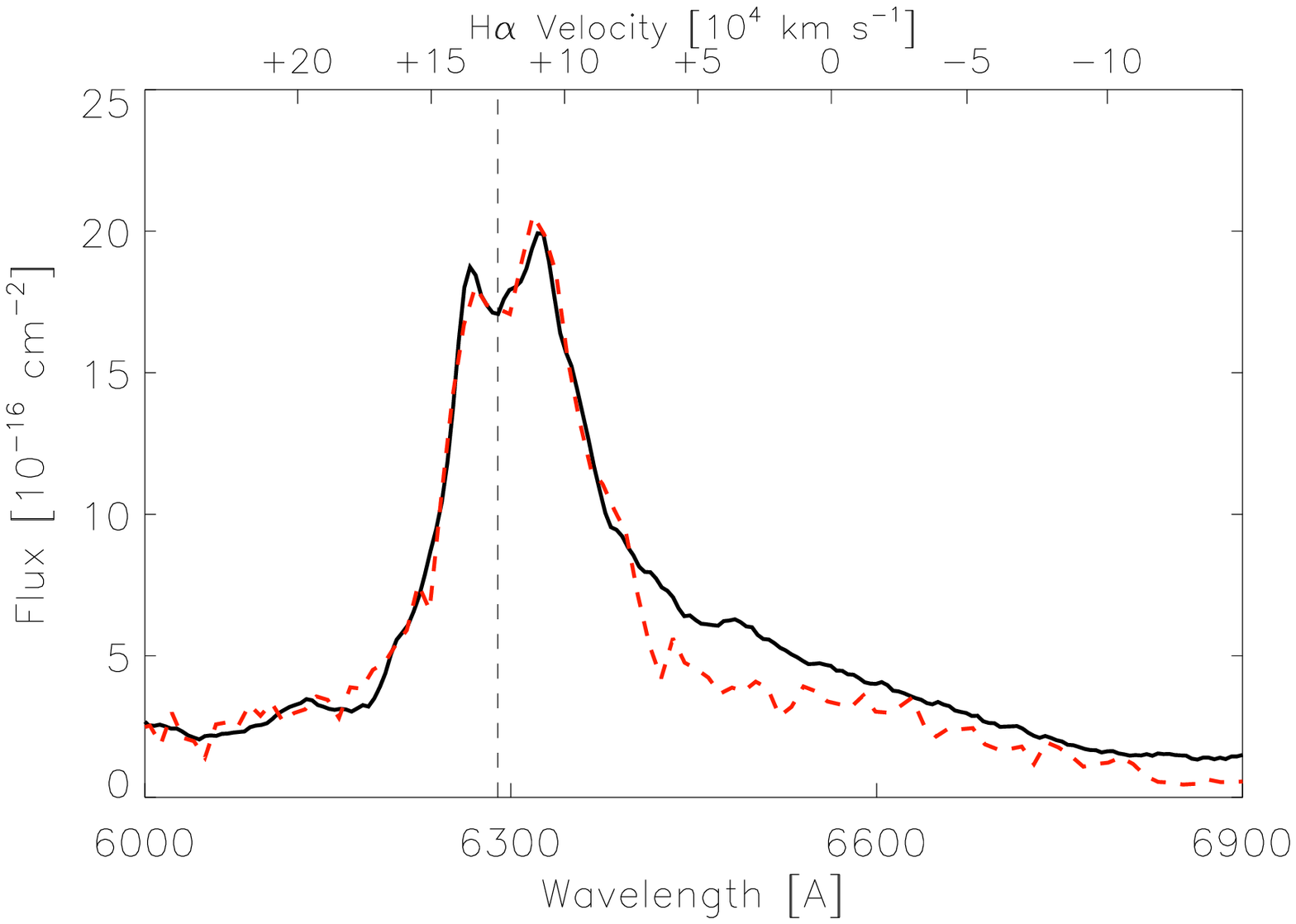}
\end{center}
\caption{Optical spectrum of SN 2008ax at 149 days after
  explosion. Observations are
  shown in black while the synthetic flux is shown by the dashed red
  line. The synthetic flux is obtained using the three-dimensional model described in the
  text. Although the He {\sc i} IR lines are shaped by the torus-like He distribution, all other lines are single peaked, since the
  heavy elements are
  concentrated in the core. The (double) peak of the [O {\sc
    i}] $\lambda\lambda$ 6300, 6363 doublet profile is caused by
  H$\alpha$ absorption between $\sim$ 12000 and 12500 km
  s$^{-1}$. The H$\alpha$ absorption minimum is indicated by the vertical dashed line. There seems to be continuum
    flux, especially between
    7000 and 10000 \AA , which is
not reproduced. This is because of the epoch (149
days), which is too early for a strictly nebular treatment.}
\label{pap5sn08ax149}
\end{figure}

\begin{table*}
 \centering
 \begin{minipage}{140mm}
  \caption{SN 2008ax. (A) Best fit
    three-dimensional model (B) All $^{56}$Ni confined below 2800 km
    s$^{-1}$; inconsistent with nebular Fe-group line
    observations. Hydrogen is not included in the total mass and
    kinetic energy
    estimate in both models but should be of the order of 0.1 M$_\odot$ and
    10$^{50}$ ergs.}
  \begin{tabular}{ccccccccc}
  \hline
   & He & C & O & Ca & Ni & M$_\mathrm{Tot}$ & E$_\mathrm{K}$\\
    & M$_\odot$ & M$_\odot$  & M$_\odot$ &  M$_\odot$ &
  M$_\odot$ &  M$_\odot$ & 10$^{51}$ ergs  \\
 \hline
  A  & 2.0 & 0.09 & 0.51 & 0.005 & 0.10 & 2.7 &
  $\ge$ 0.9 \\
  B  & 2.7 & 0.07 & 0.40 & 0.002 & 0.07 & 3.3 &
  $\ge$ 1.2\\
\hline
\end{tabular}
\end{minipage}
\end{table*}

\begin{table*}
 \centering
 \begin{minipage}{140mm}
  \caption{Properties of SN 2008ax obtained by different
    methods: (A) Multi-dimensional nebular modelling (B) one-zone light-curve calculation (C) semi-analytical
    light-curve modelling (D) numerical light-curve modelling using the
    radiation transport code STELLA on a SN IIb explosion
    model (E) light-curve comparison of SN 2008ax and SN 1993J.\newline $^*$Please note
    that (C), (D) and (E) are computed for $E(B-V)$ = 0.3 mag instead of
    0.4 mag. In rough approximation a comparison can be made by multiplying
    all quantities of (C), (D) and (E) by a factor of 1.15. These values
    are given in brackets.}
  \begin{tabular}{ccccccccc}
  \hline
    & $\mu$ & $E(B - V)$& M$_\mathrm{^{56}Ni}$  &M$_\mathrm{Tot,ej}$ & E$_\mathrm{K}$ & Reference\\
    &  mag & mag & M$_\odot$ &M$_\odot$ & 10$^{51}$ergs & \\
 \hline
 A &  29.92 $\pm$ 0.29 & 0.4 $\pm$ 0.1 & 0.10$^{+0.05}_{-0.03}$ & 2.8$^{+1.2}_{-0.9}$ & 1.0$^{+1.1}_{-0.3}$ & This work \\
 B  & 29.92 $\pm$ 0.29 & 0.4 $\pm$ 0.1 & 0.07$-$0.15   &4.6 $\pm$ 2.5  &  1.0$-$10  & [Tau10] \\
 C  & 29.92            & 0.3           & $\sim$ 0.06$^*$ [0.07] &
 $\sim$ 2.9$^*$ [3.3]    &  $\sim$ 0.5$^*$ [0.6] & \citet{Roming09}\\
 D  &  29.92            & 0.3           & $\sim$ 0.11$^*$ [0.13] &
 $\sim$ 2.3$^*$
 [2.7]         &  $\sim$ 1.5$^*$ [1.7]  & \citet{Tsvetkov09}\\
 E  & 29.92 $\pm$ 0.29 & 0.3 & 0.07$^*$ $-$ 0.11$^*$ [0.08 - 0.13] &
 3$^*$ $-$ 6$^*$ [3 - 7] & $\sim$ 1.0$^*$ [1.2] & \citet{Pastorello08}\\
\hline
\end{tabular}
\end{minipage}
\end{table*}

Nebular
spectra of SN 2008ax are available at 131, 149, 266, 280 and 359 days after
explosion. The spectrum at 131 days covers the range between 9000 and 25000 \AA
. The other spectra cover a range between 5000 and 10000 \AA . For the modelling we use a distance
modulus of $\mu$ = 29.92 $\pm$ 0.29 mag and an extinction of $E(B-V)$
= 0.4
$\pm$ 0.1 mag \citep[see][]{Taubenberger10}[Tau10]. Also see
\citet{Pastorello08,Chornock10} for observations and discussion of SN 2008ax.

There is observational evidence for asymmetry in SN 2008ax. The double-peaked
profiles of the He {\sc i} IR 10830 and 20587 \AA\ lines can be
interpreted as a torus-shaped distribution of helium [Tau10;\citet{Chornock10}]. In addition, the
blue wing of these lines is stronger, which can be interpreted as an
asymmetry along the line of sight, but may also be a scattering
effect. We place the observer in the equatorial plane, which is necessary to
produce the double-peaked He profiles in this model. 

The choice of this geometry (see Figure \ref{pap5sn08axpic} for
illustration) is motivated by the observations but is probably not unique. Different geometries
may reproduce acceptable fits to the observations as well.

In contrast to the He IR lines (see Figure \ref{pap5sn08ax131}), most other lines are single
peaked (see Figure \ref{pap5sn08ax149}). This is expected, since heavy elements are concentrated in
the core. An
exception is the [O {\sc i}] $\lambda\lambda$ 6300, 6363 doublet. While we agree that the
profile of these lines may be shaped by geometry [Tau10] in other types of
CC-SNe
\citep[e.g.][]{Mazzali05,Maeda08,Modjaz08,Taubenberger09,Maurer09}, we think that in SNe IIb H$\alpha$ line scattering is
responsible for the splitting of the [O {\sc i}] $\lambda\lambda$ 6300,
6363 doublet (also see Section \ref{other} for other SNe IIb). 
 
The velocity of the H$\alpha$ absorption minimum
saturates at $\sim$ 12500 km s$^{-1}$ about 40 days after
explosion. This is usually
interpreted as the lower boundary of hydrogen. Since hydrogen in lower
layers may be strongly ionised (see Appendix~A), this estimate is somewhat
uncertain. It may rather be the lower boundary of H {\sc i} but not
of H {\sc ii}. 

We can reproduce the double-peaked profile of
the [O {\sc i}] $\lambda\lambda$ 6300,
6363 doublet very well placing less than 0.1 M$_\odot$ of hydrogen
between 12000 and 12500 km s$^{-1}$ (see Figure
\ref{pap5sn08ax149}). An exact estimate of the H mass is not possible since shock
interaction or clumping may be important in this region. This can explain why other lines,
like [O {\sc i}] 5577 \AA\ are single-peaked. This is
discussed in more detail in Section \ref{disc}.

Our model consists of 128 angular and 8 radial cells (spherical
geometry; 1024 cells in total). The radial cells have outer boundaries at 2000, 2800, 4500, 6000, 6600, 9500, 12000 and 12500
km s$^{-1}$. This choice relates to the physical
properties of our model for SN 2008ax. 

Below 2000 km s$^{-1}$ the ejecta are spherically symmetric. This
innermost part of the SN is dominated by $^{56}$Ni, oxygen and
calcium. The total mass and mass fractions in this zone are quite
uncertain, but their contribution to the global properties of the SN is
small. No He is present ($<$ 10$^{-4}$ M$_\odot$).

The region between 2000 and 2800 km s$^{-1}$ is similar to the
innermost part, but some cells, preferentially in the equatorial plane,
contain small fractions of He ($\sim$ 10$^{-3}$ M$_\odot$). The bulk of the oxygen, calcium and carbon of SN
2008ax is located in this region. In contrast to calcium, carbon is
probably not present below 2000 km s$^{-1}$. The confinement of C to a thin low
velocity layer is in agreement with theoretical predictions \citep[e.g.][]{Nomoto93}.

The region between 2800 and 4500 km s$^{-1}$ is dominated by oxygen
and $^{56}$Ni in some cells and by He in others. This separation
improves the reproduction of the observations. On the
one hand He emission can be observed down to 2000 km s$^{-1}$ and the
He IR line ratio suggests that there is no strong mixing with heavy
elements (which decreases the electron temperature and influences the
strength and
ratio of the He {\sc i} 10830 and 20587 \AA\ lines). On
the other hand, the iron lines observed are much broader than 2800 km
s$^{-1}$. Placing all the $^{56}$Ni mass necessary to reproduce the
observed flux below 2800 km s$^{-1}$ causes too narrow and too strong
Fe emission lines. A good compromise is found by allowing a separation
of the $^{56}$Ni and He-rich ejecta. Large-scale structures of burned and
unburned material are not unexpected depending on the explosion
scenario. The He rich cells contain some fraction of carbon, oxygen,
sodium and calcium (in total $\sim$ 30\%). These elements are strongly
excited by the energy absorbed by the He layer.

The material above 4500
km s$^{-1}$ is dominated by helium, which constitutes most of the mass
of SN 2008ax. The He rich layer reaches out to at least $\sim$ 9500 km s$^{-1}$.
The composition and density of the zone between 9500 and 12000 km
s$^{-1}$ is not clear. The absorption profile of the He {\sc i} IR
lines suggests that there is no He {\sc i} 20587 \AA\ line scattering above $\sim$ 10000
km s$^{-1}$ (see Figure \ref{pap5sn08ax131}). An upper limit to the He
mass in this region is $\sim$ 0.05 M$_\odot$. The observation of the [O {\sc i}] $\lambda\lambda$ 6300,
6363 doublet shows that there is no strong H$\alpha$ line scattering
below 12000 km s$^{-1}$ (see Figure \ref{pap5sn08ax149}). A possible solution may be that there is some
 He (of the order of 0.1 M$_\odot$) in this region, but mixed with small
 fractions of neutral elements, like H {\sc i}. In this case  continuum destruction of He
 {\sc i} 584 \AA\ photons could reduce the optical depth of the He
 {\sc i} 20587 \AA\ line significantly \citep{Chugai87,Li95}. This effect strongly depends on the composition of the He
layer. However, since the mass in this
region is expected to be low, the effect on our total mass estimate
would presumably be less than 10\%. The estimate of the kinetic energy
could be influenced more strongly.

The 266, 280 and 359 day H$\alpha$ observations cannot be reproduced using any reasonable
amount of hydrogen (of the order of 0.1 M$_\odot$).  It seems likely that some
additional mechanism of H$\alpha$ emission becomes important between
150 and 200 days after explosion, as in SN 1993J [Hou96]. In Section
\ref{clumping} we propose a mechanism, based on mixing and
clumping of hydrogen and helium. Shock interaction is also a
possibility and is discussed in Section~\ref{disc}.

The O {\sc i} 7774 \AA\ line is reproduced well at 266 days and later
but is too weak at 149 days after explosion, which is expected
\citep{Maurer10} since there is no clumping of the ejecta in our model.

The He {\sc i} 10830 \AA\ line may contain some contribution from Si
and S lines. However, in SN 2008ax this contribution is weak.
It seems likely that He {\sc i} 20583 \AA\ and possibly
He {\sc i} 10830 \AA\ are influenced by line scattering. Both lines
have optical depths larger than one around 130 days after
explosion. However, it is not clear if there is enough flux that can
be scattered. Our nebular code produces almost no emission on the blue
side of both lines. However, at least for He {\sc i} 20587
\AA\ this is in conflict with the observations, which show some background flux around 20000 \AA . Since the spectrum
at 131 days is probably not completely nebular, it is not surprising that
certain features of the spectrum are not reproduced well (also see the
background flux in the optical spectrum at 149 days). 

To handle this
problem we introduce some artificial background flux between 19000 and
22000 \AA\ at the flux level of the observations. This allows the He
{\sc i} 20587 \AA\ line to increase by line scattering. However, since
this background is completely artificial, one can not expect that the
modelling is accurate. 

The masses estimated for He, C, O, Ca and $^{56}$Ni are shown in Table
1. The uncertainties on these estimates should be of the order of
(several) 10\%. Other elements carry larger
uncertainties. We also compute the total mass and
kinetic energy, excluding hydrogen. Our best-fit
three-dimensional model is listed Table 1, row (A). Since some SN IIb
models predict a confinement of $^{56}$Ni to quite low velocities
\citep{Nomoto93}, we  additionally show
an estimate obtained by placing all $^{56}$Ni below 2800 km s$^{-1}$ (B),
which is, however, in conflict with the Fe-group emission line
observations.

A comparison of the total mass of models (A) and (B) shows that the uncertainties owing to
the distribution of $^{56}$Ni is about half a solar
mass. However, since model (B) stands in clear contradiction to the
nebular Fe-group emission line observations model (A) has to be
preferred. There are also uncertainties from He {\sc i} background
scattering, clumping and from the
atomic data. Additionally, we may underestimate the mass in the outer
regions. 

Owing to all the uncertainties described above, we
estimate a total mass between 2.2 and 3.2
M$_\odot$ and a $^{56}$Ni mass between 0.10 and 0.12
M$_\odot$.  A lower
limit on the kinetic energy is 0.9 $\cdot 10^{51}$
ergs (for a total mass of 2.7 M$_\odot$). Since small amounts of high velocity H and He could increase
the total kinetic energy considerably we estimate a total kinetic
energy between 0.7 and 1.7~$\cdot$~$10^{51}$~ergs  (for a total mass
between 2.2 and 3.2 M$_\odot$).

In addition there is some uncertainty ($\sim$ 40\%) owing to extinction (we used $E(B - V)$
= 0.4 $\pm$ 0.1 mag) and distance (9.6 $\pm$ 0.3 Mpc). Since these
uncertainties would influence the total mass and the $^{56}$Ni mass
estimate simultaneously we assume that the uncertainty owing to distance and
extinction is $\sim$ 20\% on each of these quantities. Thus we
estimate (including 0.1 M$_\odot$ of hydrogen with a kinetic energy of
10$^{50}$ ergs) a total mass of
2.8$^{+1.2}_{-0.9}$ M$_\odot$, a total $^{56}$Ni mass of
0.10$^{+0.05}_{-0.03}$ and a total kinetic energy 1.0$^{+1.1}_{-0.3}
\cdot 10^{51}$ ergs. 

In Table 2 we compare our estimates to results from other
groups. [Tau10] obtain estimates from one-zone-modelling
of the
light curve. They estimate a total mass of $\sim$ 4.6 M$_\odot$, a
kinetic energy of 6 $\cdot$ 10$^{51}$ ergs and a
$^{56}$Ni mass of 0.1 M$_\odot$. The estimates of the total  mass and
kinetic energy are larger than found in
this work. The uncertainties
estimated by [Tau10] are large and the results
agree within these uncertainties. \citet{Roming09} model the light-curve
of SN 2008ax using a combination of an analytical light-curve model and
a Monte-Carlo routine. Our results are roughly consistent, however the
kinetic energy estimated by \citet{Roming09} seems to be too low in
general. This has also been found by [Tau10].

A comparison to \citet{Tsvetkov09} shows that our results agree rather
well. Since the extinction was estimated to be lower in the work
of \citet{Tsvetkov09} we multiply their results by a factor of 1.15 for
comparison. This gives 0.127 M$_\odot$ of $^{56}$Ni, a total mass of
2.7 M$_\odot$ and a kinetic energy of 1.7$\cdot$ 10$^{51}$ ergs. We
may underestimate the kinetic energy but our results
agree within the errors.

Our estimates for the $^{56}$Ni and total mass, as well as for the
kinetic energy are also consistent with the results of \citet{Pastorello08}.

\section{Other SNe of Type IIb}
\label{other}
Nebular models for SNe 1993J [Hou96], 2001ig \citep{Silverman09}, 2003bg
\citep{Mazzali09} and 2007Y \citep{Stritzinger09} exist but
a treatment of helium was not possible, since there are no IR nebular
spectra ([Hou96] included He in their
analysis but had no IR observations for comparison). An exact treatment of hydrogen is also difficult since it
is observed as H$\alpha$, which may be influenced by scattering, clumping and shock interaction.

Therefore, we restrict our analysis to the question: can H$\alpha$ be
powered by radioactive energy deposition, as described by [Hou96]? The answer to this question is less model dependent than
estimating element masses or distributions. At late epochs ($>
200$ days) the SN ejecta are illuminated homogeneously by
$\gamma$-radiation (since the $\gamma$-optical depth is low) and
asymmetries in the $^{56}$Ni or hydrogen distribution are not very
important. In addition we can obtain some constraints on the H and He
distribution from early and late-time line-width observations. The maximum H mass is
restricted by light curve observations and should be less than one solar
mass \citep{Nomoto93,Utrobin94,Woosley94}. 

Early time observations show that all SNe of our sample have
absorption minima of H$\alpha$ between 13000 and 20000 km s$^{-1}$
during the first days after explosion. The velocity of these
absorption minima decreases rapidly in the following tens of days and
bottoms out at $\sim$ 10000 $-$ 13000 km s$^{-1}$ in all SNe
around 30 $-$ 40
days after explosion. This saturation behaviour is usually interpreted
as the lower boundary of the hydrogen layer. However, there may be
some hydrogen at lower velocities if it is ionised completely in the
early phase of the SN (see Appendix~A). We show that there is
an intriguing link between the velocity of the early time H$\alpha$
absorption minimum and the profile of the [O {\sc i}] $\lambda\lambda$
6300, 6363 doublet for all SNe of our sample (see Table 3).

Whether a double-peaked [O {\sc i}] $\lambda\lambda$
6300, 6363 doublet is
formed or not, depends on
the velocity of the optically thick H$\alpha$ . The optimum H$\alpha$
velocity to produce a double-peaked O [{\sc i}] profile is  $\sim$
12000 km s$^{-1}$ since then the absorption
minimum of H$\alpha$ for 6300 \AA\ lies right at the centre of the
[O {\sc i}] $\lambda$ 6300 line profile, which corresponds to the centre of the
SN (bulk of oxygen).

\begin{table}
\caption{The velocity of the absorption minimum of H$\alpha$  at
  $\sim$ 40 days after the explosion ($v_\mathrm{phot,H\alpha}$) taken from the literature and the
  velocity of the H$\alpha$ derived from the [O {\sc
      i}] $\lambda\lambda$ 6300, 6363 doublet profile in this
  work ($v_\mathrm{neb,H\alpha}$). A physical connection between both velocities seems
  likely, confirming that the profile of the [O {\sc
      i}] $\lambda\lambda$ 6300, 6363 doublet is strongly influenced
  by H$\alpha$ absorption.}
  \begin{tabular}{cccccccccc}
  \hline
  SN  &  $v_\mathrm{phot,H\alpha}$ & $v_\mathrm{neb,H\alpha}$  & M$_\mathrm{H}$ & Reference \\
   	  & km s$^{-1}$& km s$^{-1}$ & M$_\odot$ &    \\
 \hline
  1993J   & $\sim$ 10000 & $\le$ 11000 & $\sim$ 0.2 & \citet{Woosley94} \\
 2001ig   & $\sim$ 13500 & $\sim$ 13200    & $-$    & \citet{Maund07} \\
 2003bg   & $\sim$ 13000 & $\sim$ 12800 & $\ge$ 0.05 & \citet{Mazzali09}\\
  2007Y   & $\sim$ 10000 & $\le$ 11000 &  $-$ & [Tau10]\\
  2008ax  & $\sim$ 12500 & $\sim$ 12200 &  $-$ & [Tau10]  \\
\hline
\end{tabular}
\end{table}

\subsection{SN 1993J}
\label{sn93j}

\begin{figure} 
\begin{center}
\includegraphics[width=8.5cm, clip]{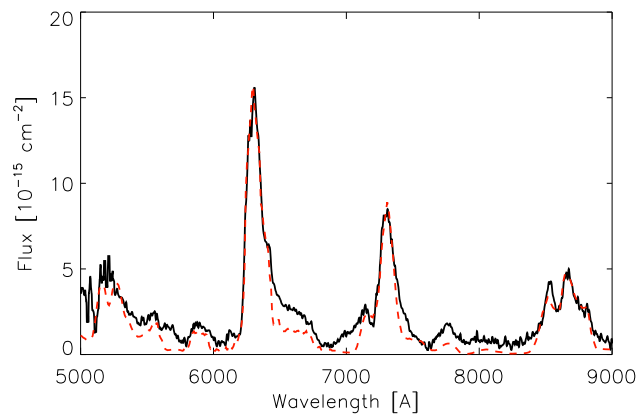}
\end{center}
\caption{Optical spectrum of SN 1993J at 206 days after explosion (black line). The synthetic flux (red dashed line) is
  produced using 0.2 M$_\odot$ of hydrogen distributed between 7000
  and 10000 km s$^{-1}$. H$\alpha$ is not reproduced with sufficient
  strength. The [O {\sc i}] $\lambda\lambda$ 6300, 6363 doublet is
  single peaked, since hydrogen is located below 11000~km~s$^{-1}$.}
\label{pap5sn08ax206}
\end{figure}

Nebular spectra of SN 1993J have been
modelled before, investigating the formation of the H$\alpha$ line in detail
[Hou96]. We repeat this analysis to test whether our
results are in agreement with previous findings. We have 
spectra at 118, 172, 206, 237, 256, 300 and 363 days after
explosion. While the
118 day spectrum is not strictly nebular, all the later ones are. The spectra cover a range between 4000 and 10000 \AA\ but
no IR observations are available. For the modelling we use a distance
modulus of $\mu$ = 27.72 mag and an extinction of $E(B-V)$ = 0.18 mag [Hou96].

[Hou96; see their Figure 1] have shown that a model consisting of a heavy element core (0
$-$ 3400 km s$^{-1}$), a He layer (3400 $-$ 7800 km s$^{-1}$) and a
hydrogen dominated layer ($ >$ 7800 km s$^{-1}$) containing $\sim$ 25\% of
He can produce synthetic spectra consistent with the nebular
observations of SN 1993J (except H$\alpha$ at late epochs). 

Using a model similar to the one presented by [Hou96] we can reproduce the evolution of the heavy element
lines of SN 1993J at all epochs between
118 and 363 days (e.g. see Figure \ref{pap5sn08ax206}). The hydrogen
mass of SN 1993J was estimated to be 0.2 M$_\odot$
\citep{Woosley94,Houck96}. Using this hydrogen mass we come to the same conclusions as [Hou96]. The synthetic H$\alpha$ flux and the observations
become inconsistent between 150 and 200 days after explosion. This is usually
interpreted as the time of the transition
to a shock interaction dominated phase \citep[e.g.][]{Patat95,Houck96}. 

The [O {\sc i}] $\lambda\lambda$ 6300, 6363 line in SN 1993J
is not double-peaked. As we have shown for SN 2008ax, the double-peaked
oxygen profile is possibly the result of H$\alpha$ absorption. Since
the optically thick H$\alpha$ of SN 1993J has a velocity $<$ 11000 km
s$^{-1}$ (see Table 3), the oxygen line is not split by the H$\alpha$ absorption
minimum. This is discussed in more detail in Section \ref{disc}. 

\subsection{SNe 2001ig \& 2003bg}
\label{sn01ig}

\begin{figure} 
\begin{center}
\includegraphics[width=8.5cm, clip]{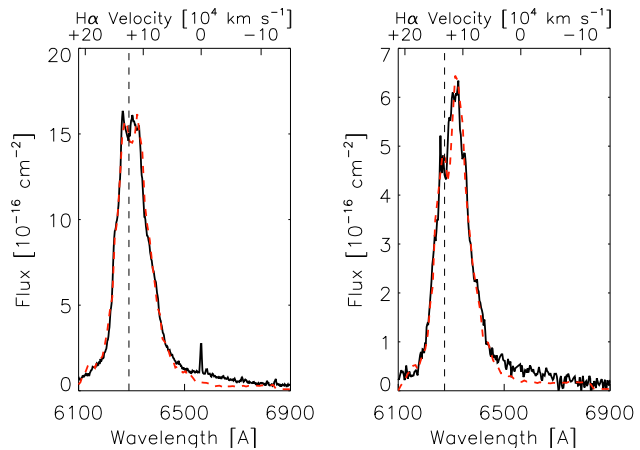}
\end{center}
\caption{Left: Optical spectrum of SN 2001ig at 340 days after
  explosion. Right: Optical spectrum of SN 2003bg at
301 days after explosion. The band shown (6100 to 6900 \AA ) is
dominated by the [O {\sc i}] $\lambda\lambda$ 6300, 6363
doublet. The observations
are shown by the black lines. The red dashed lines show the synthetic
flux obtained using $\sim$ 0.3 M$_\odot$ of hydrogen and a
\textbf{one-dimensional} model. Hydrogen causes
some weak line scattering of the oxygen-dominated flux around 6300 \AA
, creating the double peaked profile of the oxygen doublet, but
does not provide enough H$\alpha$ flux to explain the
observations around 6560 \AA . The double peaked oxygen profile is not caused by geometry, but by
H$\alpha$ absorption around 13200~km~s$^{-1}$ (SN 2001ig)
and 12800~km~s$^{-1}$ (SN 2003bg). The H$\alpha$ absorption minima are
indicated by the vertical dashed lines.}
\label{pap5sn01ig03bg}
\end{figure}

For SN 2003bg there is a H mass
estimate of $\sim$ 0.05 M$_\odot$ obtained from early-time absorption
modelling \citep{Mazzali09}, while there is no estimate of the H mass of SN 2001ig.

We have nebular spectra of SN 2001ig at 256, 309 and 340 days after
explosion. The spectra cover a range between 4000 and 10000 \AA . For the modelling we use a distance
modulus of $\mu$ = 30.5 mag and an extinction of $E(B-V)$ = 0.011 mag \citep{Silverman09}.

For SN 2003bg we have nebular
spectra at 264 and 301 days after
explosion. The spectra cover a range between 4000 and 10000 \AA . For the modelling we use a distance
modulus of $\mu$ = 31.68 mag and an extinction of $E(B-V)$ = 0.02 mag \citep{Mazzali09}.

Although the nebular H$\alpha$ flux is weak in both SNe, we find that
it cannot be reproduced using reasonable amounts of hydrogen.
In contrast to the H$\alpha$ emission, the absorption can be reproduced
well. The [O {\sc i}] $\lambda\lambda$ 6300, 6363 doublet is often used
to investigate asymmetries of CC-SNe cores. However, in the case of SNe
IIb there is strong evidence that the profile of the oxygen
line is not shaped by geometry alone but also by absorption (see
Section 3). To reproduce the
oxygen profiles of SN 2001ig and 2003bg we have to place  $\sim$ 0.3 M$_\odot$ of
hydrogen between 13000 and 13500 km s$^{-1}$ (SN 2001ig) and 12600 and
13000 km s$^{-1}$ (SN 2003bg) which is roughly consistent with
the early-time absorption minimum of H$\alpha$ (see Table 3) of these SNe. 

Although we can reproduce
the absorption features well, it seems unlikely that 0.3 M$_\odot$ of
hydrogen are necessary to reproduce the observations if there is
some additional mechanism of H$\alpha$ excitation (e.g. shock
interaction). This may be expected,
since H$\alpha$ emission is underestimated in our
models.

\subsection{SNe 2007Y}
\label{sn07Y}

\begin{figure} 
\begin{center}
\includegraphics[width=8.5cm, clip]{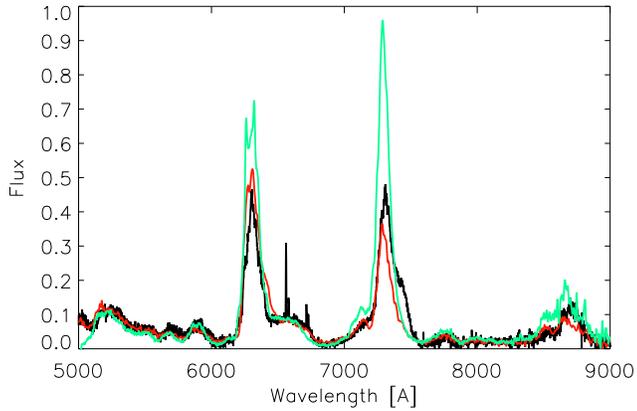}
\end{center}
\caption{Optical spectra of SNe 1993J (256 days; red line), 2007Y (248
days; black line) and 2008ax (266 days; green line). The spectra are
scaled by arbitrary constants. Apart from the [O~{\sc
    i}]~$\lambda\lambda$ 6300, 6363 doublet and the [Ca {\sc ii}] emission the
spectra agree extremely well. The ratio of H$\alpha$ to oxygen
emission is strongest in SN 2007Y, which is surprising since this SN was
classified as SN Ib, while SNe 1993J and 2008ax were classified
as Type IIb. SN 2007Y has the strongest nebular ratio of
H$\alpha$ to oxygen (and total) flux ever detected in a stripped CC-SN. The [O {\sc i}] $\lambda\lambda$ 6300, 6363 doublet is single-peaked, which is expected if the bulk of hydrogen is located below 11000 km s$^{-1}$.}
\label{pap5sn93J07Y08ax}
\end{figure}

Although H$\alpha$ absorption was detected in its early spectra, SN
2007Y was classified as a SN Ib \citep{Stritzinger09}. While these
H$\alpha$ features would be sufficient to classify SN 2007Y as a Type IIb, there is more evidence for the
presence of hydrogen since there is strong
H$\alpha$ emission in the nebular phase. \citet{Stritzinger09}
argue that the emission is caused by shock interaction, but \citet{Chevalier10} claim
that the circumstellar density of SN 2007Y is too weak to produce the
H$\alpha$ luminosities observed before 300 days after explosion. 

The nebular spectra cover
a range between 4000 and 10000 \AA . For the modelling we use a
distance modulus of $\mu$ = 31.43 mag and an extinction of $E(B-V)$ =
0.112 mag \citep{Stritzinger09}.
There is no IR nebular spectrum.

The H$\alpha$ flux can be reproduced neither at 248 nor at 288 days using reasonable amounts of hydrogen.
We therefore conclude that the
H$\alpha$ flux of SN 2007Y cannot result from radioactive energy
deposition as described by [Hou96] even if it was a SN IIb and that some other mechanism
is needed. To avoid a double-peaked profile of the [O {\sc i}] $\lambda\lambda$
6300, 6363 doublet, the hydrogen must be concentrated below 11000 km s$^{-1}$.
This is in perfect agreement with observations of
the early-time H$\alpha$ absorption minimum (see Table 3).

SN 2007Y has the highest H$\alpha$ to [O {\sc
    i}] $\lambda\lambda$ 6300, 6360 doublet flux ratio (and also total
flux) of all five SNe (see Figure
\ref{pap5sn93J07Y08ax}). Unless a serious amount of
hydrogen has been accreted from a thick circumstellar wind, SN 2007Y
must have had a significant fraction of H in its outer layers at the
time of explosion. In addition, H$\alpha$ is observed at $\sim$ 15000
km s$^{-1}$ during the first days after explosion. 

Therefore, SN
2007Y is most likely a SN of Type IIb similar to SN 1993J and SN 2008ax. 
Then it is important to understand why the early
time H$\alpha$ absorption was weak \citep{Stritzinger09}. A simple
explanation may be that the hydrogen of SN 2007Y is more mixed with helium than the hydrogen of SN 1993J and 2008ax. In this
case hydrogen could be ionised more strongly at early times, which means that there is hydrogen but no H {\sc i}. Thorough mixing of hydrogen and helium could also explain the
strong, low-velocity H$\alpha$ emission observed at late epochs (see
Section \ref{clumping}).
This is discussed in more detail in Section \ref{disc}.

\section{An alternative to shock interaction}
\label{clumping}

\begin{figure} 
\begin{center}
\includegraphics[width=8.5cm, clip]{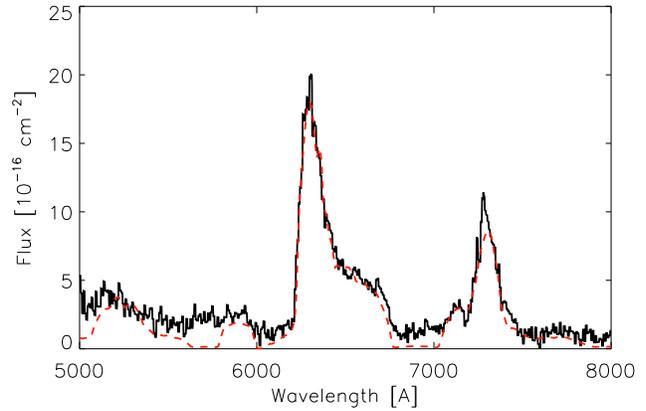}
\end{center}
\caption{Optical spectrum of SN 1993J at 363 days after
  explosion (black line). The synthetic flux (red dashed line) was produced by a one-dimensional model
  containing 2 M$_\odot$ of He and 0.2 M$_\odot$ of
  hydrogen. H and He are mixed  and distributed out to $\sim$ 10000 km
  s$^{-1}$. A clumping factor of the He layer of $\zeta = 200$ was
  used. As expected the [O
    {\sc i}] $\lambda\lambda$ 6300, 6363 doublet is single peaked.}
\label{pap5sn93J363c}
\end{figure}

\begin{figure} 
\begin{center}
\includegraphics[width=8.5cm, clip]{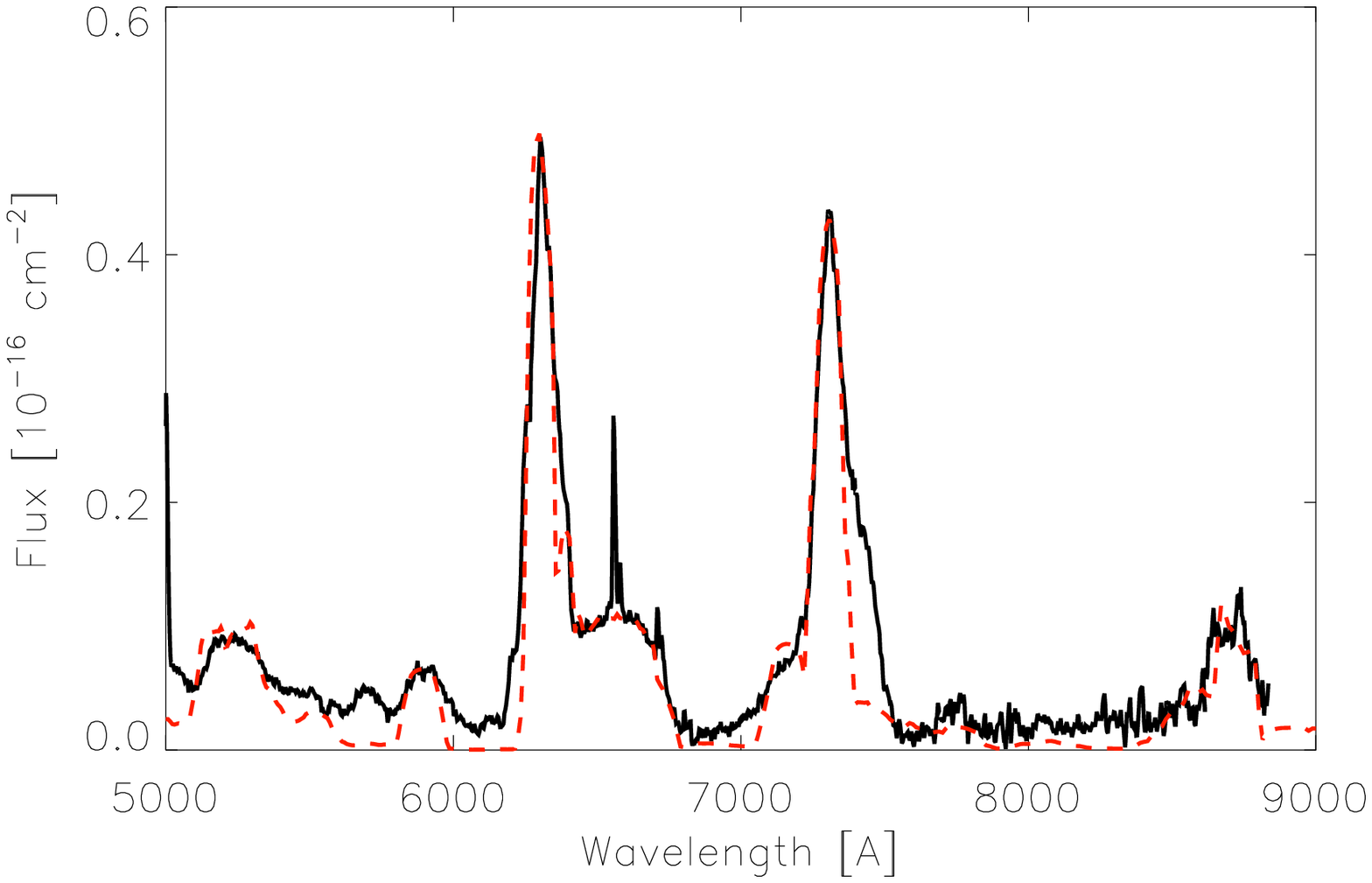}
\end{center}
\caption{Optical spectrum of SN 2007Y at 288 days after
  explosion (black line). The synthetic flux (red dashed line) was produced by a one-dimensional model
  containing 1.5 M$_\odot$ of He and 0.1 M$_\odot$ of
  hydrogen. H and He are mixed  and distributed out to $\sim$ 10500 km
  s$^{-1}$. A clumping factor of the He layer of $\zeta = 100$ was
  used. As expected the [O
    {\sc i}] $\lambda\lambda$ 6300, 6363 doublet is single peaked.}
\label{pap5sn07Y288c}
\end{figure}

\begin{figure} 
\begin{center}
\includegraphics[width=8.5cm, clip]{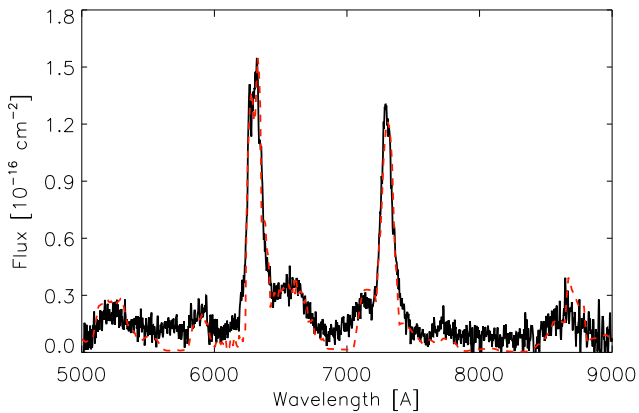}
\end{center}
\caption{Optical spectrum of SN 2008ax at 359 days after explosion
  (black line). The synthetic flux (red, dashed line) was
  produced using a one-dimensional model. A clumping factor of the He layer
  of $\zeta \sim 100$ was used. While the H$\alpha$ emission is
  produced by tiny fractions of H in the He layer below 12000 km
  s$^{-1}$, the double-peaked profile of the [O
    {\sc i}] $\lambda\lambda$ 6300, 6363 doublet is again produced by
  H$\alpha$ absorption between 12000 and 12500~km~s$^{-1}$ (also see
  Figure \ref{pap5sn08ax149}).} 
\label{pap5sn08ax359c}
\end{figure}

To our knowledge there are only three stripped CC-SNe (SNe 1993J,
2007Y, 2008ax) which show
strong, box-shaped H$\alpha$ emission in their nebular phase. It is
commonly assumed \citep[e.g.][]{Filippenko94,Patat95, Houck96} that the box-shaped H$\alpha$ emission of SN 1993J
is caused by shock interaction. 

[Hou96] have shown that radioactive energy deposition is
too weak to produce strong H$\alpha$ emission at late epochs. Their
analytical estimates do not depend on the SN 1993J model explicitly
and should be valid for any SN IIb with moderate amounts of H ($\sim$
0.1 M$_\odot$). One may therefore conclude that the same mechanism
operating in SN 1993J is also at work in SNe 2007Y and
2008ax. However, for SN 2008ax there are several problems
connected to the shock-interaction interpretation of its H$\alpha$
nebular emission [Tau10]. In SN 2007Y shock
interaction is not expected at all \citep{Chevalier10}.

In this section we present an alternative mechanism for powering late
nebular H$\alpha$ emission. This mechanism may explain
how late-time H$\alpha$ emission can be powered by radioactive decay
without any need for shock interaction, solving the problems mentioned
above. This mechanism is simple but needs special conditions to operate, which could
explain why strong nebular H$\alpha$ emission is rare in stripped
CC-SNe (this may also be explained by shock interaction).

We assume that a small fraction of hydrogen (of the order of 1\%) is mixed
into a helium-dominated layer. The degree of H
ionisation will lie between 1 and almost 100\%
depending on the ionisation state of helium, since He recombination
can ionise hydrogen strongly (see Appendix~A). There is a sharp
transition between complete and almost no ionisation of
hydrogen, if the hydrogen mass is comparable to the mass of He {\sc
  ii}. In such a situation, when the degree of ionisation of helium
decreases with time, as it does naturally, large amounts of neutral hydrogen
can be produced in regions where no H {\sc i} had been present at
earlier epochs (this is explained in more detail in Appendix~A). Therefore, the H {\sc i} fraction in the He dominated layer can
increase from approximately zero to, say, 2\% of the total mass within several hundred
days. Up to this epoch an observer would not even know about hydrogen in
these regions since H {\sc ii} cannot be observed, except by extremely weak
recombination radiation.

It is important to note that the energy deposited in the He-rich
regions (at least in SN 2008ax, but probably also in SNe 1993J and 2007Y)
is $\sim$ 10 times larger than the energy emitted in H$\alpha$. This
means that if only 10$\%$ of the deposited energy was transformed
into H$\alpha$, shock interaction would not be needed to explain the
late-time H$\alpha$ emission.

Under typical nebular conditions $\sim$ 70$\%$ of the energy deposited in a He rich layer by radioactive
decay is stored in thermal electrons. It is therefore clear that
emission from this region is dominated by thermal electron
collisional excitation and not by recombination. If heavy elements (C, O, Na, etc.) are mixed into the He layer in sufficient fractions,
they can be excited by thermal electrons and cool the gas
efficiently. In this case H$\alpha$ is predominantly produced by
recombination, which is much too weak
to explain the late time H$\alpha$ flux [Hou96; they neglected
  thermal electron excitation of hydrogen]. Since this scenario would
produce high-velocity optical lines (e.g. oxygen or sodium), which are
not observed, helium cannot be mixed with heavy
elements strongly (at least in SN 2008ax; see Section
\ref{sn08ax}).

If the gas consists of hydrogen and helium only, or if heavier
elements are singly ionised to almost 100$\%$, which would be the case if the
fraction of these elements was lower than the fraction of He {\sc
  ii}, thermal electrons cannot cool effectively until they reach high
enough temperatures to excite hydrogen ($\sim$ 10000K ; see Appendix~A). In this
scenario 99$\%$ of the thermal electron energy is radiated away by
Ly$\alpha$ emission and therefore lost to the UV. About 1\% is converted
into H$\alpha$, which is too small by a factor of $\sim$ 10 to explain the
H$\alpha$ observations.

At this point clumping could become important. We use the symbol
$\zeta$ for the clumping factor, which is defined as the inverse of
the filling factor [see \citet{Maurer10} for more details]. A clumping factor
$\zeta$ means that the density is increased by a factor
of $\zeta$ locally, while the global density remains constant. 

A helium layer
containing small amounts of hydrogen can emit much more than 1\% of
its thermal electron energy in H$\alpha$ if clumping is strong, since
clumping increases the excitation of H {\sc i} to n = 2 \& 3, and the
self-absorption optical depth of the 
H {\sc i} ground state transitions. 

Summarising, this means that, if hydrogen and helium are
mixed in suitable fractions and clumped strongly, radioactive energy
deposition can power H$\alpha$ completely without any additional source
of energy. We present some estimates here, which support this idea. 

Since the excitation potential of
hydrogen is $\sim$~10~eV lower than that of
helium, the thermal electron collisional excitation rates of hydrogen are
approximately exp(10eV/kT) $\sim$ 10$^{5-3}$ times (at 10000 $-$ 20000K) higher than those of helium. Unless there are $\sim$
10$^{5-3}$ times more He {\sc i} than H {\sc i} atoms $\sim$ 70$\%$ of
the total
energy deposited in helium will be radiated away by hydrogen and He IR
lines (the other
30\% will go into non-thermal electron excitation and recombination radiation, mainly UV).

It is clear that, as long as H is ionised completely ($\gg 99$\%), H$\alpha$
emission and scattering are extremely weak. There is no H$\alpha$
scattering since there is no Ly$\alpha$ self-absorption if there is no
H {\sc i}, which means that the H {\sc i} n = 2 level is
depopulated strongly. Moreover, there is almost no H$\alpha$ emission, since
only $\sim$ 20$\%$ of the deposited energy go into ionisation. This energy
is mainly emitted by He {\sc i} recombination radiation and Ly$\alpha$,
which means that the fraction of deposited energy emitted by H$\alpha$
is less than one percent. Thermal electrons cool predominantly by
exciting He {\sc i}, since the temperature increases to more than
15000K.

We derive a rough estimate for the dependence of H$\alpha$ emission on
clumping. We consider effective recombination into the hydrogen levels n = 2
and 3, thermal electron (de-) excitation between the ground state and
the levels n = 2 and n = 3, and radiative transitions from H {\sc i} n
= 3 to n = 2 and from n = 2 to n = 1. We neglect radiative transitions
from H {\sc i} n = 3 to n = 1, since this rate is small owing to
self-absorption. We also neglect thermal excitation from H {\sc i} to n = 2 to n = 3,
which may contribute to the population of the H {\sc i} n = 3 level
but is less important than direct excitation from the ground state.
We only
consider the situation where enough H {\sc i} is present to cool the
gas efficiently ($\frac{n_\mathrm{H[I]}}{n_\mathrm{He}} > 0.1$\%), which means that the H {\sc i} ground state transitions are
highly self-absorbed.

To calculate the formation of H$\alpha$ the population
of the n = 2 state of H {\sc i} has to be estimated
\begin{equation}
\begin{split}
n_\mathrm{H[I],2} & \sim \frac{C_{12}\zeta [+ If_2]}{A_{21}\tau_{21}^{-1}\zeta^{-1} +
  C_{21}\zeta}n_\mathrm{H[I],1}, \qquad \tau_{21} \gg 1\\
& \propto \zeta^2, \qquad \zeta\sim 1, C_{12} > I\\
& \propto \zeta^0, \qquad \zeta\gg 1\\
\end{split}
\end{equation}
where $C_\mathrm{ij}$ are the thermal electron collisional rates
between the n = i and j states of H {\sc i} \citep[e.g.][]{Callaway94},
I is the total ionisation rate (non-thermal electrons and
photo-ionisation; see Appendix~A), $f_n$ are
the fractions of recombining electrons, cascading into the H~{\sc i} n~states
and 
\begin{equation}
\tau_\mathrm{21} = \frac{\lambda_\mathrm{Ly\alpha}^3tg_2A_{21}}{8\pi
  g_1}n_\mathrm{H[I],1}, \quad n_\mathrm{H[I],1} \gg n_\mathrm{H[I],2}
\end{equation}
is the Sobolev optical depth of the H {\sc i} ground state, with
$\lambda_\mathrm{Ly\alpha} \sim$ 1216 \AA\ the wavelength of Ly$\alpha$,
$g_\mathrm{n}$ the statistical weights of the n states of H {\sc
  i} and $A_{21}$ the radiative rate of Ly$\alpha$. Depending
on the number density of the H {\sc i} ground state the n = 2 state can be
thermally populated ($C_{21}\zeta \gg \frac{A_{21}}{\tau_{21}\zeta}$) at moderate clumping
factors ($\zeta \sim 10$) already.
The H$\alpha$ optical depth is given by
\begin{equation}
\begin{split}
\tau_\mathrm{H\alpha} &  \sim \frac{\lambda_\mathrm{H\alpha}^3tg_3A_{32}}{8\pi
  g_2}n_\mathrm{H[I],2}, \quad n_\mathrm{H[I],2} \gg n_\mathrm{H[I],3}\\
& \propto \zeta^2, \qquad \zeta \sim 1\\
& \propto \zeta^0, \qquad \zeta \gg 1 \\
\end{split}
\end{equation}
where $\lambda_\mathrm{H\alpha} \sim$ 6563 \AA\ is the wavelength of
H$\alpha$ and $A_{32}$ is the radiative rate of H$\alpha$. $\tau_\mathrm{H\alpha}$ increases with clumping
at low clumping factors, but then saturates as soon as the de-population
of the H {\sc i} n = 2 level is dominated by collisional de-excitation.
In rough approximation the ratio of the H$\alpha$ to the Ly$\alpha$ luminosity is given by
\begin{equation}
\begin{split}
\frac{\mathrm{H}\alpha}{\mathrm{Ly}\alpha} & \sim
\frac{E_\mathrm{H\alpha}}{E_\mathrm{Ly\alpha}}\frac{n_\mathrm{H[I],3}}{n_\mathrm{H[I],2}}\frac{A^\prime_\mathrm{32}}{A^\prime_\mathrm{21}}\\
& \sim
\frac{E_\mathrm{H\alpha}}{E_\mathrm{Ly\alpha}}\frac{C_{13}\zeta  [
    +
  If_3]}{C_{12}\zeta [ + If_2]}\\
& \times \Big(1 + \frac{C_{21}\tau_{21}\zeta^2}{A_{21}}\Big)\Big(1 +
\frac{C_{31}\tau_\mathrm{H\alpha}\zeta^2}{A_{32}[1 - \exp(-\tau_\mathrm{H\alpha}\zeta)]}\Big)^{-1}\\
& \sim 0.03 \cdot\exp\Big(-\frac{E_\mathrm{H\alpha}}{kT}\Big),
\quad\zeta\sim 1, C_\mathrm{1j} > I\\
& \sim \mathrm{F}(T, n_\mathrm{e}, n_\mathrm{H}, n_\mathrm{H[II]},\zeta), \quad \zeta\gg 1\\
\end{split}
\label{Lalpha}
\end{equation}
where $A^\prime_{ij}$ denotes the self-absorbed radiative transition
rates between the n = i and j states of H~{\sc i}.

Clumping can increase the relative strength of H$\alpha$ significantly. High clumping factors ($\zeta \gg 1$) increase the
H$\alpha$ optical depth less effectively than they do the H$\alpha$
luminosity, while moderate clumping factors increase both.  An
estimate for the time-dependent density $n_\mathrm{H[I]}$
is available (see Appendix~A). 

A numerical computation of the processes described above can be
performed using our nebular code. In Figures \ref{pap5sn93J363c},
\ref{pap5sn07Y288c} and \ref{pap5sn08ax359c} we show models of SN 1993J (363 days), 2007Y (288
days) and SN 2008ax (359 days) using the mechanism described
above (radioactive decay energy; mixing of H and He; strong clumping). Since the
late H$\alpha$ emission of SNe 2001ig and 2003bg is much weaker it
is clear that the late H$\alpha$ emission of these SNe can be
reproduced in a similar way. For SN 1993J
we use the He mass estimate of \citet{Woosley94}. For SN 2007Y we
assume that the He mass is a bit smaller than in SN 1993J. For SN 2008ax the He
mass is estimated in Section \ref{sn08ax}. For SN 1993J we
use a helium mass of 2 M$_\odot$, a hydrogen mass of 0.2 M$_\odot$ and a clumping factor $\zeta \sim 200$, for SN
2007Y we use a helium mass of 1.5 M$_\odot$, a hydrogen mass of 0.1
M$_\odot$ and a clumping factor $\zeta \sim 100$ and for SN 2008ax we use a helium
mass of 2 M$_\odot$, a hydrogen mass of 0.2 M$_\odot$ and a
clumping factor of $\zeta \sim 100$. Most of the hydrogen is concentrated
in a thin shell at high velocities, while tiny fractions are mixed into
the lower velocity helium. It is difficult to decide whether
clumping, the helium or the
hydrogen mass should be increased to obtain a reproduction of the
spectra. Several combinations are possible.  Additionally, the outer
region could be influenced by shock interaction. Therefore it is not
possible to derive a hydrogen or helium mass estimate from this procedure.  
It seems however clear that small amounts of hydrogen are sufficient
to reproduce late H$\alpha$ emission (which is usually not possible [Hou96]),
if there is strong clumping ($\zeta \sim 100$). 

It is important to note that we could not reproduce the temporal
evolution of the H$\alpha$ line using this mechanism with our nebular code. We can find models
that reproduce the observations at any epoch, but no model describing
the observations at all epochs consistently. This is discussed in Section \ref{disc}.

\section{Discussion}
\label{disc}

\subsection{Late H$\alpha$ emission} 

We have investigated the formation of H$\alpha$ in the
nebular phase of five SNe IIb (re-classifying SN 2007Y as SN IIb). We find that radioactive energy from $^{56}$Ni decay is
not sufficient to power the H$\alpha$ emission observed at late
nebular epochs ($>$ 200 days), as long as the energy deposited in
the He layer cannot be tapped. This finding is consistent with
the work of [Hou96] for SN 1993J, but raises the question of which mechanism causes the observed H$\alpha$ flux.

In Section \ref{clumping} we have
shown that a combination of mixing and strong clumping could solve this problem. It is
not clear whether such high clumping factors are realistic (however
see e.g. \citet{Kozma98}). Furthermore, we could
not reproduce the temporal evolution of H$\alpha$ using this
mechanism. This may be because of several
reasons. First, there is a degeneracy between the hydrogen and the
helium mass and clumping and it is not clear which combination has to
be chosen. Secondly, as can be seen from
Equation \ref{Lalpha}, the strength of H$\alpha$ is very sensitive to
the electron density and temperature at clumping values of the order 100. In addition,
$n_\mathrm{H[I,II]}$ is extremely sensitive to mixing and clumping. The
composition has to be known at the one percent level or better. Our
nebular code is expected to be less accurate. We do not treat UV-radiation transport explicitly, the
ionisation of hydrogen by helium is treated in approximation only and
in general there are uncertainties owing to the atomic data and owing to
the incomplete description of the physical scenario. Asymmetries of
the ejecta can further complicate the problem.

Since the radioactive scenario appears to have problems, it may be worth considering
the most common interpretation, i.e. that late SN IIb H$\alpha$ emission is the
result of shock interaction. An extensive discussion
of several shock interaction scenarios for SN 2008ax can be found in
[Tau10]. However, as pointed out by [Tau10], there are several problems
within all these scenarios. Shock interaction has
difficulties explaining why H$\alpha$ emission is observed at low
velocities in SN 2007Y and SN 2008ax (see [Tau10]). Moreover, there are
contradictions between X-ray and H$\alpha$ observations, if both are
interpreted as shock interaction (see SNe 2003bg, 2007Y). These problems may arise from an incomplete
understanding of the influence of the shock radiation on the ejecta or
from an inaccurate treatment of the
physical parameters determining the shock properties (e.g. the SN
outer density structure). 

The shock interaction model can explain why strong H$\alpha$ emission
at late times is rare in SN IIb and has the advantage that it may explain a
flattening of the late H$\alpha$ light curve, which is observed in SN
1993J and possibly in SN 2008ax. However, this has never been shown
quantitatively. Nor has it been shown that shock interaction
can reproduce late H$\alpha$ emission in detail in any SN IIb at all.

The mechanism presented in Section \ref{clumping} can explain the formation of low velocity H$\alpha$
in SNe 2007Y and 2008ax, which is observed. It can further explain the absence of
high velocity H$\alpha$ emission, again consistent with the observations, since in this
scenario the H$\alpha$ emission traces the He distribution. It can
explain why SN 2007Y, has strong H$\alpha$ emission, although
\citet{Chevalier10} claim that there is extremely weak shock
interaction in SN 2007Y. It can explain why SN 2003bg has no strong
H$\alpha$ emission, although \citet{Chevalier10} claim that there is
strong shock interaction in SN 2003bg. Since the model needs some fine
tuning it can explain why strong late time H$\alpha$ is
rare in SNe IIb.

There may
well be some other mechanism (not clumping; still mixing
of H and He), which emits in H$\alpha$ energy absorbed by the He layer.
If it was possible to use $\sim$
10$\%$ of the energy of a massive He shell to
excite H$\alpha$ without
producing any other strong optical lines, the late-time
H$\alpha$ emission could be explained by radioactive energy deposition alone.

\subsection{Nebular line profiles of SNe IIb}
Usually, double-peaked line profiles are interpreted as toroidal
ejecta distributions. While we agree with this interpretation
in general (e.g. for SN Ib/c; see also Section \ref{sn08ax}) we think that the situation may be different in SNe IIb. We have shown that, taking the 40 day (after
explosion) H$\alpha$ absorption minimum velocity as lower boundary of
the bulk of hydrogen (as it is usually done) we can explain the peak
profile of the [O {\sc i}] $\lambda\lambda$ 6300, 6363 doublet for all
five SNe of our sample consistently by H$\alpha$ absorption.

Since H$\alpha$ absorption causes a split of the [O {\sc i}]
$\lambda\lambda$ 6300, 6363 doublet if it is located around
12000~km~s$^{-1}$, 
and since the position of H$\alpha$ influences the position of this
split, we can infer the radial distribution of H {\sc i} from
fitting the [O {\sc i}]
$\lambda\lambda$ 6300, 6363 doublet profile. 
For all five SNe of our sample
the H$\alpha$ velocity measured with this method agrees very well
with the early-time H$\alpha$ minimum velocity (see Table 3). It seems unlikely that
this is coincidence. 

The outer H$\alpha$ region could be optically thick even at late
epochs (e.g. excited by shock interaction),  which could
explain why the profile of the [O {\sc i}] $\lambda\lambda$ 6300, 6363
doublet does not change its shape significantly at late
epochs. For example, the reverse shock of SN 2008ax could be located around 12000
km s$^{-1}$ at 359 days after the explosion [Tau10].

\subsection{SN 2008ax}
We have derived a three-dimensional model of SN 2008ax. Two kinds of
asymmetries may be observed. First the He IR lines are double-peaked,
which is most easily explained by a torus shaped He
distribution. Secondly, the blue sides of the He lines appear to
be stronger than the red sides. 

We find a total ejecta mass of $\sim$ 3 M$_\odot$, containing about 0.1
M$_\odot$ of $^{56}$Ni and expanding with a kinetic
energy of $\sim 10^{51}$ ergs, which is consistent
with findings from light curve and early-time
H$\alpha$ absorption modelling (Table 2; also see [Tau10]). We find a helium
mass of $\sim$ 2 M$_\odot$ and derive the abundances of heavier
elements such as carbon or oxygen (Table 1). 

While most of the heavy element mass is confined to velocities below
4500 km s$^{-1}$, helium is predominantly found between 4500 and
probably 12000 km s$^{-1}$. The outer boundary is uncertain.
Some helium is present below 4500 km s$^{-1}$.
We find that a separation of helium and heavy elements, at least
between 2000 and 4500 km s$^{-1}$, improves the reproduction of the He
IR lines.

Since our nebular code does not reproduce the continuum flux observed
around 20000 \AA , we added some artificial background to allow He {\sc
  i} 20587 \AA\ line scattering. There seems to be no He {\sc i}
20587 \AA\ line scattering above $\sim$ 10000 km s$^{-1}$, while the
He {\sc i} 10830 \AA\ line would not be in conflict with higher
helium velocities. The observations of the He {\sc i} 20587 \AA\ line
place an upper limit on the He mass between 9500 and 12000 km s$^{-1}$
of $\sim$ 0.05 M$_\odot$. It may well be that continuum destruction (e.g. by hydrogen), which
could reduce the  He {\sc
  i} 20587 \AA\ optical depth, is important. In this case the He mass
in this region could be larger. 

We can compare our results to SN IIb models of
\citet{Nomoto93} and \citet{Woosley94}. The \citet{Nomoto93} models [also see Hou96]
predict a strong confinement of $^{56}$Ni to low velocities ($<$ 2000
km s$^{-1}$). Other heavy elements are distributed in layers above the
radioactive core. Most of the hydrogen is confined to a thin layer
above the He layer. This He layer contains some traces of hydrogen and heavier
elements. The models of \citet{Woosley94} predict a broader distribution
of $^{56}$Ni in velocity space. Heavy elements dominate at low
velocities but are also mixed into the He layer. Helium and hydrogen
are mixed in the outer regions. 

We find that $^{56}$Ni is probably not as centrally confined as
predicted by the \citet{Nomoto93} models. On the other hand carbon
seems to be confined to a thin shell on top of the iron core in
agreement with these models. Since our core is deformed to an axis ratio
of $\sim$ 3:2 (see Figure \ref{pap5sn08axpic}), one probably cannot expect perfect agreement with the
spherical symmetric models. The distribution of heavy elements seems
rather consistent with the predictions in the core as well as in the
He layers. Our results for hydrogen are inaccurate and therefore a
comparison is difficult. 

We compare the masses of He, C, O, Ca and $^{56}$Ni of our SN
2008ax model to the 13C model of \citet{Woosley94}, which has been
used by \citet{Tsvetkov09} to reproduce the light curve of SN
2008ax. We find that our estimates of the  He, C, Ca and $^{56}$Ni mass agree
at the 10$\%$ level. The estimate of the oxygen mass is larger by a
factor of $\sim$ 2 than
predicted by the 13C model.
  
Thanks to its late time IR
observations, SN 2008ax is the first SN IIb
where the He density field can be determined from nebular
observations. Other methods, such as light-curve modelling, provide
poor information about element abundances or asphericities. Early-time
modelling can determine the properties of the outer layers more
accurately, but no information about the central region can be obtained. Therefore nebular IR observations (10000 to 22000 \AA ) of SNe Ib and IIb
are highly desirable.

\section{Summary \& Conclusion}
\label{sum}

We have derived estimates related to hydrogen and helium line
formation in the nebular phase of SNe IIb. We have shown that
hydrogen can be highly ionised if it is mixed with helium in suitable
ratios. This has probably been known before. However, it could explain why
H$\alpha$ can sometimes be observed in the nebular phase but not
at early times. This finding is of special interest for SN 2007Y. 

From our analysis it seems
likely that SN 2007Y is a SN~IIb. The lack of strong H$\alpha$
absorption around 50 days after explosion may well be explained by strong mixing of H
and He. Such mixing could also explain the formation of strong late time
H$\alpha$. Since we cannot determine the He distribution of SN
2007Y, this remains a speculation.

We have shown that radioactive energy deposition is insufficient to
produce the late-time H$\alpha$ emission in all SNe IIb of our sample,
as long as the energy of the helium layer cannot be
tapped by H$\alpha$. 

Shock interaction may be observed in all SNe of our
sample. There seem to be two
types of late H$\alpha$ emission scenarios. Strong, box-shaped (SNe 1993J, 2007Y,
2008ax) and weak (SNe 2001ig, 2003bg) H$\alpha$. The
simplest explanation would seem to be strong and weak shock interaction,
respectively. However, this is in conflict with findings of
\citet{Chevalier10}, at least for SNe 2003bg and 2007Y. It also seems
difficult to explain the low velocity observations in H$\alpha$ by shock interaction [Tau10].

We presented an alternative mechanism to shock interaction, explaining
late time H$\alpha$ emission by radioactive energy deposition. The right
combination of mixing and clumping of hydrogen and helium has been shown to
be able to reproduce the H$\alpha$ observations of all SNe of our
sample at least up to 350 days after explosion. However, there are
also some problems with this interpretation of late H$\alpha$
emission. Summarising, it is not clear how late H$\alpha$ is
formed in SNe IIb. Therefore, we were not able to derive an estimate
of the hydrogen mass for any SN of our sample.

We have shown that most likely the profile of the [O {\sc i}] $\lambda\lambda$ 6300, 6363
doublet is influenced by H$\alpha$ absorption strongly. Hydrogen
concentrations above $\sim$ 11000 km s$^{-1}$ can cause double-peaked oxygen profiles, while slower
hydrogen can not. We have shown that observations of early-time H$\alpha$
absorption minima and the corresponding oxygen line profiles are perfectly consistent
with this interpretation for all five SNe of our sample.

This scenario also explains why other lines, like Ca {\sc ii} or
[O~{\sc i}]~5577~\AA , are single peaked in all SNe of our sample.
An exception are the He lines of SN 2008ax, which show signs of asymmetry. It seems likely that at least the inner part of
SN 2008ax is asymmetric. We have obtained a three-dimensional model of the SN
2008ax envelope. For the first time the helium mass of a SN IIb has been
determined from nebular modelling. We have obtained estimates
of the total mass and kinetic energy of SN 2008ax in excellent
agreement with results from light-curve modelling. Chemical abundances
have been derived. The nebular model
of SN 2008ax provides the opportunity to compare observations and
theoretical SN IIb models with unprecedented richness of detail.

As already stated by \citet{Houck96}, nebular IR observations (10000 to
22000 \AA ) of
core-collapse SNe would be extremely useful for deriving
properties of CC-SNe such as helium abundance, total mass and kinetic
energy. It is important that both He IR lines (10830 and 20587 \AA ) are
observed simultaneously. Such observations should be obtained for any SN Ib or IIb.

\bibliography{pap5}

\section*{Appendix A}

\begin{figure} 
\begin{center}
\includegraphics[width=8.5cm, clip]{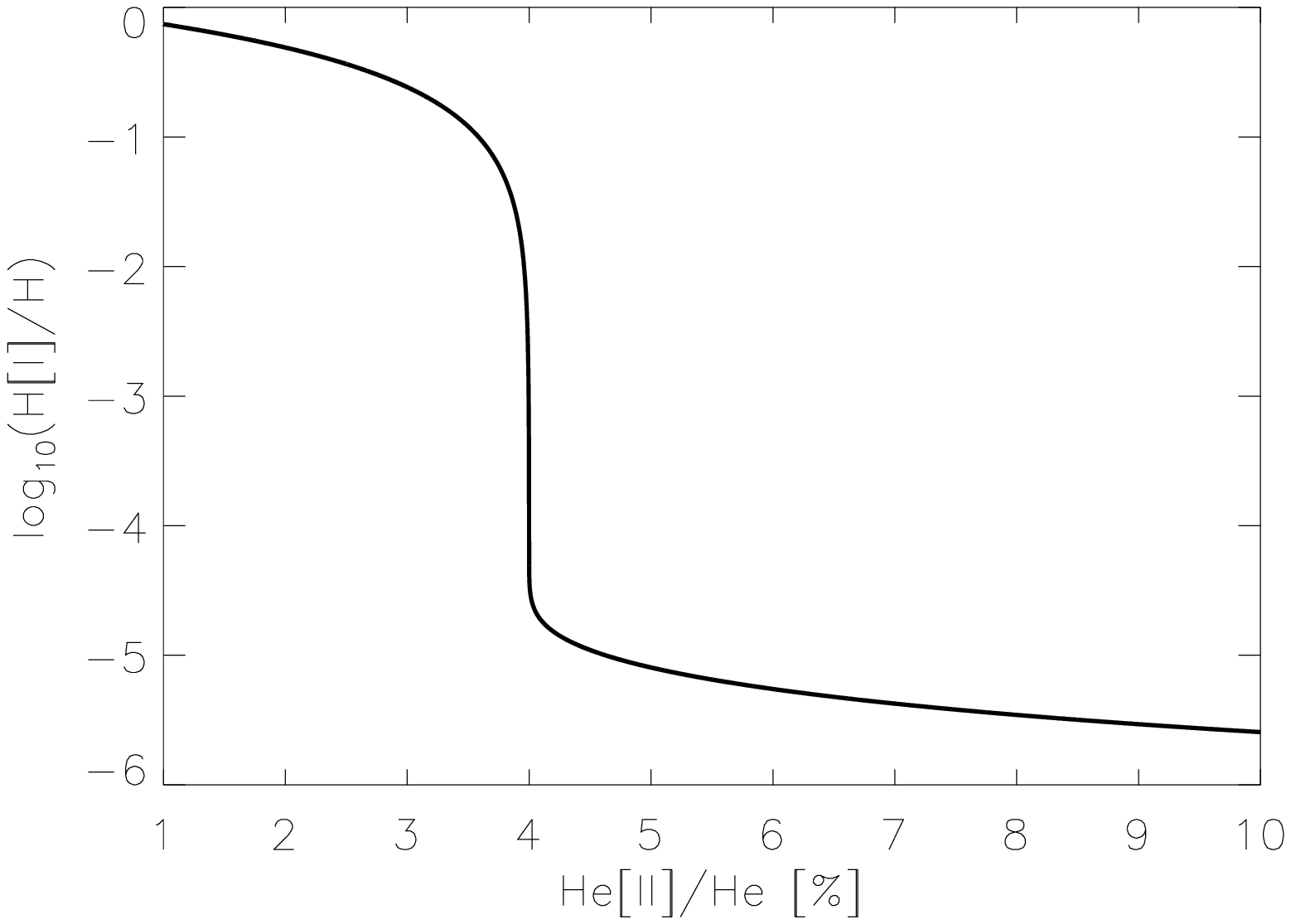}
\end{center}
\caption{The logarithmic fraction of neutral hydrogen
  $\frac{n_\mathrm{H[I]}}{n_\mathrm{H}}$ (full black line; Equation
  \ref{nhI}) as a function of the He {\sc
    ii} fraction $\frac{n_\mathrm{He[II]}}{n_\mathrm{He}}$ [\%] for a
  helium dominated layer containing 2\% of hydrogen, assuming $f$ =
  0.5 and $\xi_\mathrm{H} = 0.1$. The He density was set to 10$^8$ g
  cm$^{-3}$. There is a steep decrease of H {\sc i} as soon as the fraction of He {\sc ii}
  increases above $\frac{2}{f}$\%. Changing the degree of ionisation
  of He by 1\% only (from 4.5 to 3.5\% in this example), the fraction of H
{\sc i} increases by a factor of 10$^4$.}
\label{hhe}
\end{figure}

\begin{figure} 
\begin{center}
\includegraphics[width=8.5cm, clip]{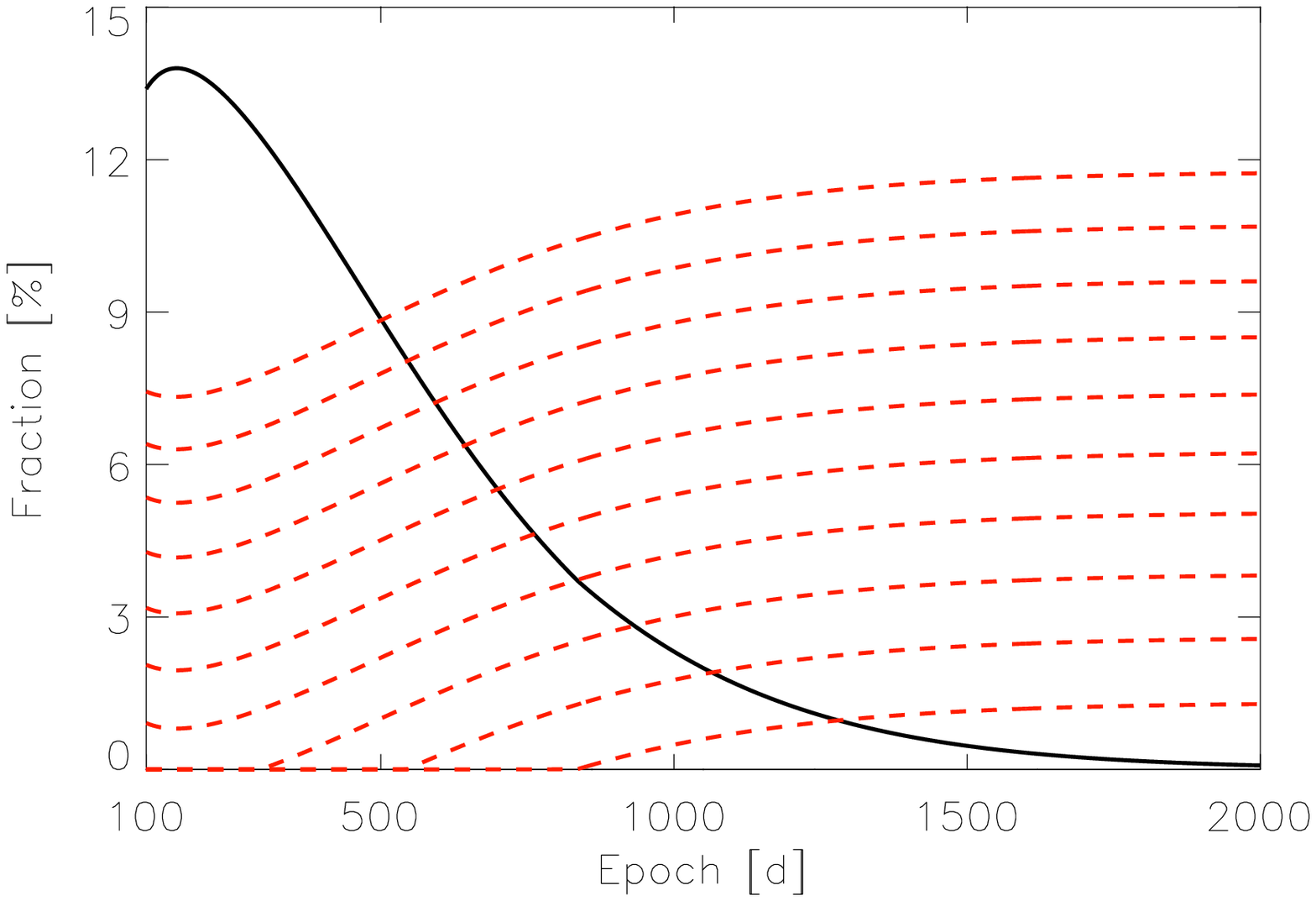}
\end{center}
\caption{Toy model of a SN IIb. The time-dependent fraction $\frac{n_\mathrm{He[II]}}{n_\mathrm{He}}$ [\%] (full black
  line; Equation \ref{nheii}) is shown for a model
  described in the text (M$_\mathrm{He}$ = 3 M$_\odot$,
  M$_\mathrm{Ni}$ = 0.1 M$_\odot$, $f = 0.5$, T = 15000K). The fraction $\frac{n_\mathrm{H[I]}}{n_\mathrm{H} + n_\mathrm{He}}$
  [\%] (dashed red line; Equation \ref{nhI}) is shown for hydrogen masses of 0.01 to 0.1
  M$_\odot$ (bottom to top). There is no clumping of the H and He
  layer. The calculation starts at 100 days since
  at earlier epochs photospheric radiation
  may be important for ionising hydrogen and helium and our estimates become invalid. Although the model is highly simplified it demonstrates that
  depending on the ratio of hydrogen and helium the H {\sc i} fraction
  can be
  approximately zero for the first 100 days and can increase rapidly
  at later epochs. This means that no H$\alpha$ would be observed at
  early epochs,
  while strong H$\alpha$ emission or scattering is possible later.}
\label{pap5nhI}
\end{figure}

\begin{figure} 
\begin{center}
\includegraphics[width=8.5cm, clip]{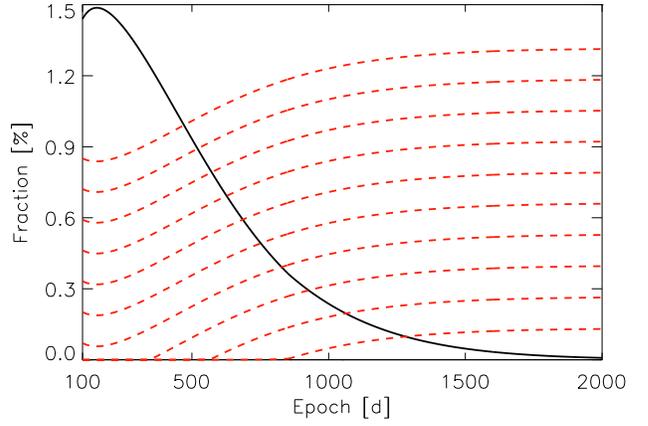}
\end{center}
\caption{Toy model of a SN IIb. The time-dependent fraction $\frac{n_\mathrm{He[II]}}{n_\mathrm{He}}$ [\%] (full black
  line; Equation \ref{nheii}) is shown for a model
  described in the text (M$_\mathrm{He}$ = 3 M$_\odot$,
  M$_\mathrm{Ni}$ = 0.1 M$_\odot$, $f = 0.5$, T = 15000K). The fraction $\frac{n_\mathrm{H[I]}}{n_\mathrm{H} + n_\mathrm{He}}$
  [\%] (dashed red line; Equation \ref{nhI}) is shown for hydrogen masses of 0.001 to 0.01
  M$_\odot$ (bottom to top). The clumping factor of the H and He layer
  is set to 100, which reduces the degree of ionisation strongly. The calculation starts at 100 days since
  at earlier epochs photospheric radiation
  may be important for ionising hydrogen and helium and our estimates become invalid. Although the model is highly simplified it demonstrates that
  depending on the ratio of hydrogen and helium the H {\sc i} fraction
  can be
  approximately zero for the first 100 days and can increase rapidly
  at later epochs. This means that no H$\alpha$ would be observed at
  early epochs,
  while strong H$\alpha$ emission or scattering is possible later.}
\label{pap5nhIc}
\end{figure}

\subsection*{Hydrogen ionisation in a He-dominated layer}

We derive some estimates for the ionisation of hydrogen by non-thermal
electrons and by the
UV-radiation emitted by helium. We assume that H is mixed with He but
not with other
elements. Therefore the treatment presented is only valid if heavy
elements are present in smaller fractions than hydrogen and
helium. However, other elements with ionisation potentials below
$\sim$ 20 eV would be affected in a way similar to hydrogen. 

About 40\% of all helium recombinations go directly to the ground state
and cause another ionisation of helium and possibly of
hydrogen. This recycling process increases the total ionisation rate
but is not important for the rest of the
appendix.
 
Electrons recombining to an excited state reach the ground state
by two-photon emission (2PE) of the 2s($^1$S) state during the nebular
phase. At early times, when the 2p levels are strongly excited by radiation, most electrons reach the ground state via the 2p levels. 

In the case of 2PE approximately 30\% of the ground state
transition radiation can ionise hydrogen \citep[the two photons are created
with energies between 0 and 20.6 eV; the chance of producing a photon with $E
> 13.6$ eV is $\sim$ 30\%;][]{Drake69} and in the case of 2p
transitions 100\%. Therefore, we assume that a
fraction $f$ (0.3 $-$ 1.0) of all He recombinations into excited states can ionise a hydrogen atom.  

The total ionisation rate
of hydrogen is then determined by non-thermal
electron ionisation and by radiative ionisation from He recombination
radiation. The non-thermal electron ionisation rates of neutral
hydrogen and helium are given by \citep[e.g.][]{Axelrod80}
\begin{equation}
Y = Y_\mathrm{H[I],He[I]} = \frac{L_\mathrm{Dep}}{N_\mathrm{Tot}W_\mathrm{H[I],He[I]}}
\end{equation}
where N$_\mathrm{Tot}$ is the total number of atoms and $W_\mathrm{H[I],He[I]}$ is the work per ion of hydrogen and
   helium, which depends on the ratio of electrons to
atoms and on the absolute atomic density (weakly). 

For simplicity we assume $W_\mathrm{H[I]}$~$=$~$W_\mathrm{He[I]}$ in this section, which will cause an error
of $\sim$ 30\% on the ionisation rates of H and He in the worst case.  

During the first few hundred days the SN is in ionisation equilibrium
\citep[e.g.][]{Axelrod80} and the ionisation balance of hydrogen can be
estimated as
\begin{equation}
Y\big(n_\mathrm{He[I]}D_\mathrm{H[I]}f + n_\mathrm{H[I]}\big) =
R_\mathrm{H[I]}\big(n_\mathrm{He[II]} + n_\mathrm{H[II]}\big)\zeta n_\mathrm{H[II]}
\end{equation}
where $D_\mathrm{H[I]}f$ is the fraction of He recombination radiation
ionising hydrogen with 
\begin{equation}
D_\mathrm{H[I]} = 1 - \exp(-\sigma_\mathrm{H}\Delta Rn_\mathrm{H[I]})
\end{equation}
where $\sigma_\mathrm{H}$ is the ionisation cross-section of H at
$\sim$ 20 eV and $\Delta$R is some characteristic width of the He shell.
The ionisation balance of Helium can be written as
\begin{equation}
Yn_\mathrm{He[I]} = R_\mathrm{He[I]}(n_\mathrm{He[II]} + n_\mathrm{H[II]})\zeta n_\mathrm{He[II]}
\end{equation}
Setting $R = R_\mathrm{H[I]} = R_\mathrm{He[I]}$ (which is a good approximation) one obtains
\begin{equation}
D_\mathrm{H[I]}n_\mathrm{He[I]}f + n_\mathrm{H[I]} =
\frac{n_\mathrm{He[I]}}{n_\mathrm{He[II]}}(n_\mathrm{H} - n_\mathrm{H[I]})
\label{nhhh} 
\end{equation}
The exact solution of Equation \ref{nhhh} is given by
\begin{equation}
\begin{split}
n_\mathrm{H[I]} & =  
\xi_\mathrm{H}^{-1}W\Bigg(\frac{n_\mathrm{He[I]}n_\mathrm{He[II]}}{n_\mathrm{He}}\xi_\mathrm{H}f\exp\Big[-\frac{n_\mathrm{He[I]}\xi_\mathrm{H}(n_\mathrm{H}
- fn_\mathrm{He[II]})}{n_\mathrm{He}}\Big]\Bigg)\\
& + \frac{n_\mathrm{He[I]}}{n_\mathrm{He}}(n_\mathrm{H} - fn_\mathrm{He[II]})
\label{nhI}
\end{split}
\end{equation}
where $W(x) \equiv
\sum_\mathrm{n=1}^\infty\frac{(-\mathrm{n})^\mathrm{n-1}}{\mathrm{n!}}x^\mathrm{n}$
is the Lambert $W$ function.
In approximation,
the deposition fraction $D_\mathrm{H[I]}$ is given by
\begin{equation*}
D_\mathrm{H[I]} = \left\{ 
\begin{array}{l l}
  1  \quad  & \mbox{$\sigma_\mathrm{H}\Delta
    Rn_\mathrm{H[I]} \gg 1$}\\
  \sigma_\mathrm{H}\Delta Rn_\mathrm{H[I]} \equiv
  \xi_\mathrm{H}n_\mathrm{H[I]}  \quad & \mbox{$\sigma_\mathrm{H}\Delta
    Rn_\mathrm{H[I]} \ll 1$}\\ \end{array} \right.
\end{equation*}
The ionisation cross section of hydrogen is $\sim$ 2 $\cdot 10^{-18}$ cm$^2$ at 20
eV and the SN radius is of the order of 10$^{16}$ cm during the
first few hundred days, which means that $\xi_\mathrm{H}$ will be of
order 0.01 $-$ 0.1 cm$^{3}$. The number density of neutral hydrogen is then given by
\begin{equation*}
\frac{n_\mathrm{H[I]}}{n_\mathrm{H}} = \left\{ 
\begin{array}{l l}
  \frac{n_\mathrm{He[I]}}{n_\mathrm{He}}\Big(1 -
  f\frac{n_\mathrm{He[II]}}{n_\mathrm{H}}\Big)  \! & \mbox{$\xi_\mathrm{H}n_\mathrm{H[I]} \gg 1$}\\
 \frac{n_\mathrm{He[I]}}{n_\mathrm{He} +
   n_\mathrm{He[I]}n_\mathrm{He[II]}f\xi_\mathrm{H}} \sim
 \big(fn_\mathrm{He[II]}\xi_\mathrm{H}\big)^{-1}  \! &
 \mbox{$\xi_\mathrm{H}n_\mathrm{H[I]} \ll 1$}\\ 
\end{array} \right. 
\label{HnH}
\end{equation*}
The degree of ionisation of hydrogen increases rapidly as soon as the hydrogen
number density becomes similar to the number density of He {\sc ii}
times $f$ (see Figure \ref{hhe}).

A factor of 10 or less in the
hydrogen abundance can make a difference of several orders in magnitude
in the degree of ionisation. Also, a small change in the ionisation
balance of He can have serious influence on the ionisation balance of
H. A simple example is given by a helium layer containing 2\% 
hydrogen and $\frac{2.5}{f}$\% He {\sc ii}. Hydrogen is ionised to $\gg
99\%$ then. If the amount of He {\sc ii} decreases to $\frac{1.5}{f}$\% (a decrease
of the ionisation fraction is expected at late epochs; see Figures
\ref{pap5nhI} and \ref{pap5nhIc}) the  H
{\sc ii} fraction decreases to less than 10\%. \textbf{Without changing the physical
conditions significantly, the fraction of H {\sc i} can change by several orders
in magnitude.} 

\subsection*{Toy model of a SN IIb}

To obtain some rough estimate of the electron temperature in a He-dominated layer containing some small fraction of hydrogen, we calculate the temperature-dependent ratio of
collisional excitation and recombination $\mathcal{R}$.
The thermal electron excitation coefficient
from n = 2 to n = 1 is given by
\begin{equation}
\begin{split}
C_\mathrm{12} & \sim 8.6 \times
10^{-6}\frac{n_\mathrm{e}}{T^{1/2}}\frac{\Omega_{21}}{g_\mathrm{1}}\exp(-\frac{E_\mathrm{Ly\alpha}}{kT})\\
& \equiv C_0n_\mathrm{e}
\end{split}
\end{equation}
where $\Omega_{21}$ is the effective collision strength of the hydrogen n = 1
to n = 2 transitions given by \citet{Scholz90} and $f_2, g_\mathrm{1}$
are defined in Section \ref{clumping}. This gives
\begin{equation}
\begin{split}
\mathcal{R}_\mathrm{} & =
\frac{C_\mathrm{12}\zeta n_\mathrm{H[I]}}{Rn_\mathrm{e}\zeta n_\mathrm{H[II]}f_2}\\
& = \frac{C_0}{Rf_2}\Big(\frac{n_\mathrm{H}}{n_\mathrm{H[I]}} -
1\Big)^{-1}\\
\end{split}
\end{equation}
While $C_0$ increases with temperature $R$ decreases and $\frac{C_0}{Rf_2}$
is approximately one at $\sim$ 11000K. Since hydrogen must
be excited by thermal collisions effectively in order to produce a
significant contribution to the total luminosity, $\mathcal{R}$ must be
larger than $\gg$ 1. Therefore the temperature must
be higher than $\sim$ 10000K, depending on the H {\sc i} fraction. 

At temperatures of $\sim$ 20000K (which are reached for low
fractions $\frac{n_\mathrm{H[I]}}{n_\mathrm{H}} < 0.1$; this implies
$n_\mathrm{H[I]} \le {n_\mathrm{He[II]}} \ll {n_\mathrm{He[I]}}$)
thermal excitation of He becomes important, the above estimate
becomes invalid and the relative importance of H {\sc i} emission
decreases as compared to He {\sc i}. Therefore, the temperature should be $\sim$ 10000 $-$
15000 K in order to allow effective H~{\sc i} emission.

We can use this temperature estimate to calculate the temporal
evolution of He {\sc ii} and therefore H {\sc i} for a toy model of the
SN ejecta. Of course this temperature
estimate is not exact, but it is sufficient to demonstrate how the
absolute fraction $\frac{n_\mathrm{H[I]}}{n_\mathrm{H} + n_\mathrm{He}}$ in a helium dominated layer
can evolve with time (see Figures
\ref{pap5nhI} and \ref{pap5nhIc}).

The deposited energy can be estimated analytically if the He density
is known. For simplicity we assume that $X$ solar
masses of He and $Y$ solar masses of $^{56}$Ni are distributed within a sphere of velocity $v$
homogeneously (of course this is in conflict with our assumption that
He is mixed with small fractions of heavy elements only, but this is
a toy model only and He and
$^{56}$Ni could be separated on small scales). At nebular epochs (say,
$>$ 100 days), when most of $^{56}$Ni has decayed to $^{56}$Co, the
deposited luminosity is given by 
\begin{equation}
L_\mathrm{Dep}(t) \sim 1.3 \times
10^{43}\exp\Big(-\frac{t [\mathrm{d}]}{111.4}\Big)\mathrm{M}_\mathrm{^{56}Ni}[\mathrm{M}_\odot] f_\mathrm{Dep}
\quad\mathrm{ergs~s}^{-1}
\end{equation}
where $f_\mathrm{Dep}$ is the deposition function, which can be
calculated analytically 
for a homogeneous sphere
\citep[e.g.][]{Axelrod80}. For three solar masses of helium and a $^{56}$Ni mass of $\sim 0.1$
M$_\odot$ distributed homogeneously in a sphere of 10000~km~s$^{-1}$
the luminosity deposited around 350~days is
$\sim$~10$^{39.5}$~ergs~s$^{-1}$ while the H$\alpha$ luminosity of SN
2008ax at this epoch is $\sim$ 10$^{38.5}$~ergs~s$^{-1}$ only [Tau10].

Since the ionisation potential of He is high and since we assumed that
it is the
predominant element, helium is mainly ionised by non-thermal
electrons
\begin{equation}
\mathcal{R}_\mathrm{He}Y_\mathrm{He}n_\mathrm{He[I]} = R_\mathrm{He}\zeta
(n_\mathrm{He[II]} +  n_\mathrm{H[II]})n_\mathrm{He[II]}
\end{equation}
where $\mathcal{R}_\mathrm{He}$ $\sim 1.5$ is the recycling fraction
of helium. This gives 
\begin{equation}
\begin{split}
\frac{n_\mathrm{He[II]}}{n_\mathrm{He}} & \sim \frac{R_\mathrm{He}\zeta
  n_\mathrm{H[II]} + \mathcal{R}_\mathrm{He}Y_\mathrm{He}}{2R_\mathrm{He}\zeta
  n_\mathrm{He}}\\
& \times \bigg(-1 + \sqrt{1 +
  4\frac{\mathcal{R}_\mathrm{He}Y_\mathrm{He}R_\mathrm{He}\zeta n_\mathrm{He} }{(R_\mathrm{He}\zeta
  n_\mathrm{H[II]} + \mathcal{R}_\mathrm{He}Y_\mathrm{He})^2}}\bigg)
\end{split}
\label{nheii}
\end{equation}
We compute Equations \ref{nhI} and \ref{nheii} for the toy model described above. We assume
that 50\% of all He recombinations into excited states can ionise a hydrogen atom ($f =
0.5$) and vary the hydrogen mass between 0.01 and 0.1 M$_\odot$. The temporal evolution of the H {\sc i} fraction is shown in Figure
\ref{pap5nhI}. Depending on the ratio of hydrogen and helium the
fractions of H {\sc i} can increase
dramatically between 100 and 1000 days. This means that H$\alpha$
scattering or emission can appear at late epochs, without any hydrogen being
detected at earlier times. In Equation \ref{nhI} and \ref{nheii} we assumed
ionisation equilibrium, which may become invalid at epochs of $\sim$ 500 days and
later. Therefore, we underestimate the He [II] fraction at very
late epochs. However, qualitatively the effect is the same, since
H {\sc i} ionisation is related to He {\sc ii} recombination even at
1000 days and later.

\end{document}